\documentclass[conference]{IEEEtran}
\IEEEoverridecommandlockouts
\usepackage{cite}
\usepackage{amsmath,amssymb,amsfonts}
\usepackage{algorithmic}
\usepackage{graphicx}
\usepackage{textcomp}
\usepackage{xcolor}
\def\BibTeX{{\rm B\kern-.05em{\sc i\kern-.025em b}\kern-.08em
    T\kern-.1667em\lower.7ex\hbox{E}\kern-.125emX}}
    
\usepackage{multirow}
\usepackage{listings}
\usepackage{subfigure}
\usepackage{soul}
\usepackage{booktabs}
\usepackage[ruled,vlined,linesnumbered]{algorithm2e}

\newcommand{\kw}[1]{{\ensuremath {\mathsf{#1}}}\xspace}
\newcommand{\TGraph}{\kw{PETGraphDB}}

\begin{document}

\title{PETGraphDB: A Property Evolution Temporal Graph Data Management System}

\author{\IEEEauthorblockN{Jinghe Song, Zongyu Zuo, Xuelian Lin, Yang Wang, and Shuai Ma*}
\IEEEauthorblockA{\textit{School of Computer Science and Engineering, Beihang University} \\
XueYuan Road \#37, Beijing, China, 100191 \\
\{songjh, zuozy, linxl, yangwangcs, mashuai\}@buaa.edu.cn}
}

\maketitle

\begin{abstract}
Temporal graphs are graphs whose nodes and edges, together with their associated properties, continuously change over time. With the development of Internet of Things (IoT) systems, a subclass of the temporal graph, i.e., Property Evolution Temporal Graph, in which the value of properties on nodes or edges changes frequently while the graph's topology barely changes, is growing rapidly. However, existing temporal graph management solutions are not oriented to the Property Evolution Temporal Graph data, which leads to highly complex data modeling and low-performance query processing of temporal graph queries. To solve these problems, we developed \TGraph, a data management system for Property Evolution Temporal Graph data. \TGraph adopts a valid-time temporal property graph data model to facilitate data modeling, supporting ACID features with transactions. To improve temporal graph query performance, we designed a space-efficient temporal property storage and a fine-granularity multi-level locking mechanism. Experimental results show that \TGraph requires, on average, only 33\% of the storage space needed by the current best data management solution. Additionally, it achieves an average of 58.8 times higher transaction throughput in HTAP workloads compared to the best current solutions and outperforms them by an average of 267 times in query latency.
% , while the amount of code required to implement these queries is approximately half of theirs.

\end{abstract}


\lstset{
% language=SQL,
basicstyle=\ttfamily\scriptsize,
breaklines=true,
aboveskip=-5pt,
belowskip=-9pt,
breakautoindent=true,
breakindent=0pt,
}

\section{Introduction}

Graphs are data structures frequently used to model and analyze complex real-world systems \cite{newman_structure_2003, costa_analyzing_2011}. When real-world systems change over time, dynamic graphs are generated, known as temporal graphs \cite{holme_temporal_2012}, whose nodes, edges, and properties on nodes or edges change over time and reflect alterations in the structure and function of real-world systems \cite{huang_tgraph:_2016,ma_fast_2020}. 
The ability of a temporal graph to incorporate temporal information makes it valuable for numerous applications and research areas, such as infectious disease spread \cite{p_peixoto_change_2018}, information dissemination \cite{sarkar_using_2019}, hot route discovery \cite{li_traffic_2007}, and graph pattern mining \cite{ma_fast_2020, temporal_motifs_2025}. 
Temporal graphs ensure a degree of flexibility that would be unattainable in static graphs, enhancing our ability to control them \cite{li_fundamental_2017}.

\begin{figure}
\centerline{\includegraphics[width=0.90\columnwidth]{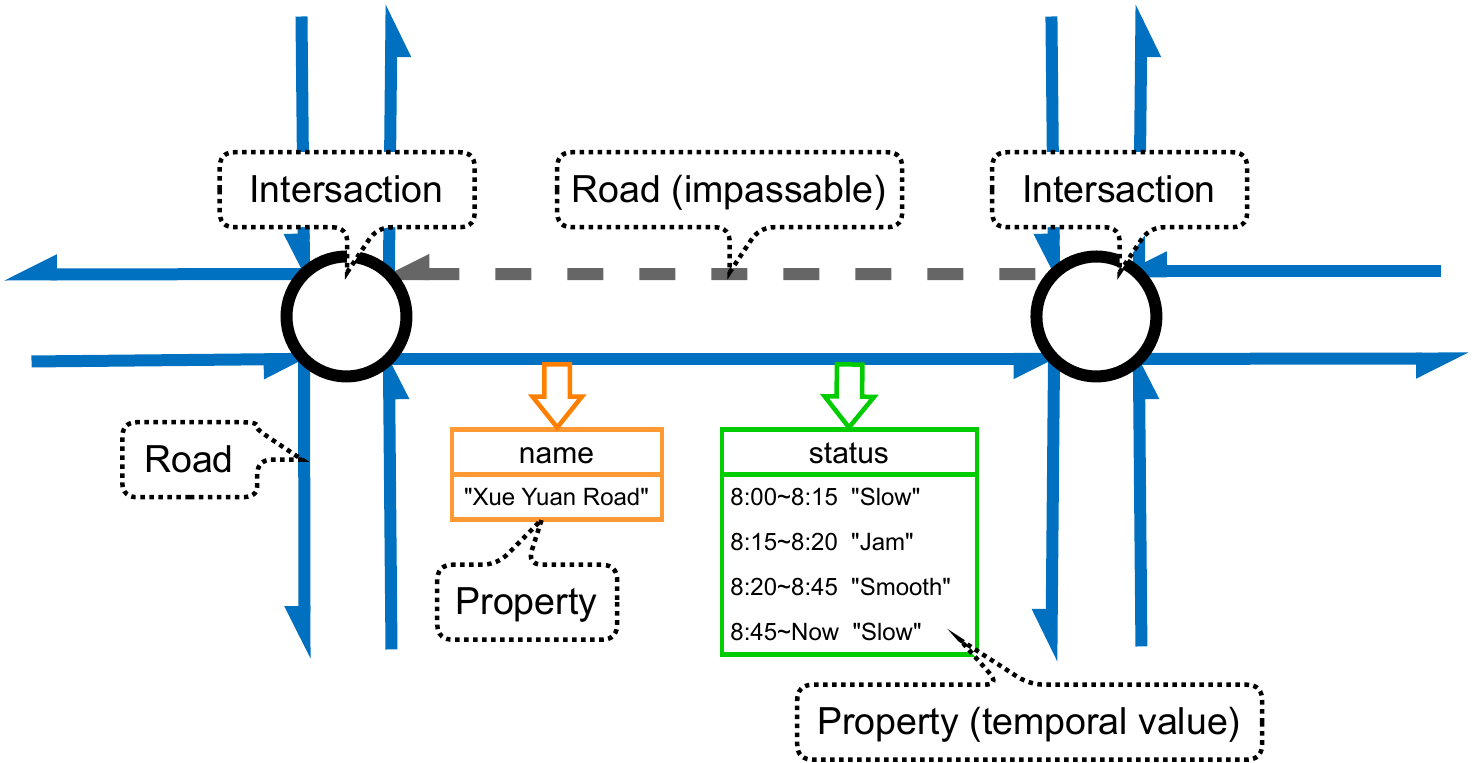}}
%\vspace{-1ex}
%\caption{Traffic network model by temporal property graph. A local subnet of the entire road network, including 2 intersections and 14 roads connected to them}
%\caption{Example of traffic network model by temporal property graph, including 2 intersections and 14 roads connected to them}
\caption{Example of traffic network model by temporal property graph, including 2 intersections connecting 14 roads}
\label{fig:traffic-model}
\vspace{-2ex}
\medskip
\end{figure}

With the rapid proliferation of IoT sensors, IoT systems are generating large volumes of a special type of temporal graphs, referred to as \underline{P}roperty \underline{E}volution \underline{T}emporal \underline{G}raph (briefly as PETG), that has a relatively stable topology but rapidly changing property values on nodes and edges. 
Fig. \ref{fig:traffic-model} illustrates an example temporal graph of a traffic system. Roads in the city are represented as directed edges (arrow lines) of the graph and intersections as nodes (black circles). The road can be passable (blue solid line) or impassable (gray dashed line). The attributes of a road are considered as properties (boxes) of the corresponding edge. Some attributes do not change with time (yellow box), such as the name and length of the road, while others may update according to time (green box), such as the jam status of roads.
In this example, the structure of the road network of a city may occasionally change when a road becomes impassable (or recovers) due to traffic control and other reasons on a daily basis, while the travel time of roads is influenced by the roads' vehicle flow and changes on a minute-by-minute basis. 
%indicating the function of the corresponding real-world system changes much more frequently than its structures.
%such as traffic systems and electronic power systems \cite{jensen_re-europe_2017}, where numerous sensors generate property changes

\begin{table*}[h!]
\centering\footnotesize
\caption{Data (access) features of some temporal graph datasets (- for no reported)}
\vspace{-1ex}
\label{tab:dataset-dist}
\begin{tabular}{@{}l|cc|cccccc@{}}
\toprule
\multirow{2}{*}{Features\textbackslash Dataset} & \multicolumn{2}{c|}{PETG} & \multicolumn{6}{c}{Other temporal graph data in related work}\\
  & Energy \cite{jensen_re-europe_2017} & Traffic \cite{temporal_motifs_2025} & Wiki\cite{massri_clock-g_2022} & Stack \cite{massri_clock-g_2022} & CitiBike\cite{massri_clock-g_2022} & Twitter\cite{miao_immortalgraph:_2015} & Weibo\cite{miao_immortalgraph:_2015} & Web\cite{miao_immortalgraph:_2015} \\
\midrule
Time span (days) & 1096 & 214 & 2320 & 2774 & 90 & 90 & 1095 & 365 \\
graph size (\# nodes/edges) & 1.5K/2.2K & 80K/110K & - & - & - & 7.5M/- & 27.7M/- & 133.6M/- \\
\# events (topology change) & 3.7K   & 190K & 8.9M & 66.0M & 2.5M & 69.1M & 4.927B & 5.641B \\
\# events (property change) & 39M  & 1.39B & 0 & 0 & 0 & 0 & 0 & 0 \\ 
\# events per day & 320K & 19.5M & 3.8K & 23.8K & 27K & 767K & 178K & 42K \\
portion (property to total) & $\geq 99.9\%$ & $\geq 99.986\%$ & \multicolumn{6}{c}{0\%} \\
property redundancy ratio & 0.05-0.5  & 0.2-0.8 & \multicolumn{6}{c}{-} \\
\bottomrule
\end{tabular}
\end{table*}

%With the rapid proliferation of IoT sensors, IoT systems based on mobile internet are generating large volumes of PETG data, which can provide new insights into analyzing the system's functional dynamics. 
PETG data provides new insights into the analysis of the functional dynamics of IoT systems. 
However, managing PETG data remains challenging due to its distinct data features and access features (see Table \ref{tab:dataset-dist}).
%that are different from other temporal graphs. 
%We collect the following features from the literature and two real-world PETG data application scenarios (the management/analytics system of a city's road transportation and an electricity network).
From a data perspective, (1) most updates are changes in property value rather than topological. In the Energy \cite{jensen_re-europe_2017} and Traffic \cite{temporal_motifs_2025} datasets, the portion of property value updates can be as large as 99.9\% of total updates, while the others are few; (2) the amount of data in the same time span is much larger (Traffic \cite{temporal_motifs_2025} dataset has an average of 19.5M updates in a day, while the Twitter\cite{miao_immortalgraph:_2015} data only has 767K updates); and (3) there is a large portion of redundant data in the property updates. In the Traffic \cite{temporal_motifs_2025} dataset, the ``jam status'' of some roads may remain ``smooth'' during the night. The redundancy ratio of some properties can be as high as 0.8 in some PETG datasets.
From the data access perspective, PETG data management should be able to (1) handle large volume data updating and supporting valid-time rather than transcation-time, (2) execute queries that dig into a range of history on a specific entity (referred to as Entity-History query)\label{ehistory-query_intro} rather than the snapshot query, and (3) keep the correctness and efficiency of concurrent read and write.

\noindent\textbf{Motivations}: 
Currently, the solutions for managing temporal graph data include the use of temporal relational databases \cite{jamkhedkar_graph_2018}, graph databases \cite{cattuto_time-varying_2013,debrouvier_model_2021,campos_towards_2016}, temporal graph databases \cite{theodorakis_aion_2024, hou_aeong_2024}, and temporal graph analysis systems \cite{massri_clock-g_2022, rost_distributed_2021, byun_chronograph_2020, khurana_storing_2016, miao_immortalgraph:_2015, labouseur_g*_2015, han_chronos:_2014, cheng_kineograph:_2012}. 
However, the lack of consideration of the features and access patterns of PETG data leads to several limitations in current solutions.
% However, these solutions do not consider the features and access patterns of PETG data mentioned above. 
% compared to the temporal graph data managed by the above solutions, 
% \st{(1) PETG data typically have much larger volumes and faster growth rates, (2) PETG data usually has non-trivial distribution patterns and input patterns, and (3) except for the snapshot query, a new class of queries that dig into a range of history on a specific entity (referred to as Entity History query) must be efficiently supported when managing such data. 
% The lack of consideration of these factors led to several limitations in the current solutions.

%\noindent\textit{Limitation 1: In-intuitive modeling}. The solutions based on non-temporal graph databases \cite{cattuto_time-varying_2013,debrouvier_model_2021} face complex modeling problems, requiring users to write logic at the application level to convert temporal graph data into a su and convert the corresponding temporal graph queries into operations supported by the underlying system, which raises the threshold for users to manage temporal graph data.

%\noindent\textit{Limitation 1: In-intuitive modeling}. The current solutions require users to write logic at the application level to convert PETG data into their native data model and convert the corresponding temporal graph queries into operations supported by the underlying system, which raises the threshold for users to manage temporal graph data. 
\noindent\textit{Limitation 1: In-intuitive modeling}. Current solutions force users to implement application-level logic that transforms PETG data into the native model of their underlying systems and rewrites temporal graph queries into the operations the systems actually support. In addition, the valid time of data in databases can not be updated by users, increasing the fexibility of managing temporal graph data.

\noindent\textit{Limitation 2: Large space usage}. IoT sensors may upload PETG data with redundancy during a relatively long period.
However, current solutions do not provide a semantically safe way to eliminate these redundancies, resulting in a significant waste of disk space.
% \st{However, the current solutions are designed for storing events rather than status, which results in significant disk space redundancy when the state changes in PETG data are fewer than event updates. This issue becomes more pronounced when the updates are automatically generated by IoT sensors at regular intervals.}

\noindent\textit{Limitation 3: High query latency}. 
% (1) Storing the PETG data in graph databases can lead to a very big average degree of nodes because the nodes/edges are split into multiple ``versioned nodes/edges'' if a property on them changes. Querying certain timeframes requires traversal of the entire history.
% (2) (Temporal) Relational databases must perform multi-table join operations for graph topology queries, which can lead to poor performance;
% \textcolor{red}{(3) When storing the PETG data in Temporal graph systems, queries may require access to a large amount of unnecessary data.}
Querying PETG data stored in current systems requires extra processes, which ``filter'' the graph topology from a mixed storage of graph topology updates and property change updates in (temporal) graph databases, or ``construct'' the graph topology through multi-table join in (temporal) relational databases. These can lead to high latency when performing temporal graph queries that frequent access topology data.

\noindent\textit{Limitation 4: Poor transaction performance}. Currently, (1) existing temporal graph analysis systems lack transaction support and cannot be used for applications requiring complete transaction management (ACID properties); and (2) locking mechanisms in existing (temporal) relational/graph database systems do not consider the features of PETG data and its queries, resulting in overly coarse-grained locks, which cause frequent mutual waiting and deadlock issues, leading to high transaction latency and low transaction success ratio.

% Data management solutions \cite{jamkhedkar_2018_nepal} based on the relational databases (with temporal extension) or time series databases (i.e. InfluxDB) are limited by their underlying implementation and only support recording the time when data is written to the database (transaction time), do not support recording the valid time of data in the real world (valid time). This limitation makes these solutions unsuitable in scenarios where the data validity time is inconsistent with the time when the data is written to the database. For example, a sensor in the traffic system encounters a network disconnect error and stores the data in its local storage. When the network recovered, the sensor re-uploaded the data to the server. Databases that only support transaction time can not handle this scenario because the occurrence time of events uploaded is earlier than the time when the events reach the database.
% % 当前只有MariaDB支持valid time的时态表功能（SQL2011标准引入）。不过并不支持在时间维度建立索引来加速访问。

\noindent\textbf{Contributions}:
In response to the above problems, this paper presents \TGraph, a temporal graph data management system oriented to Property Evolution Temporal Graphs (PETG). The main contributions of this paper are as follows:

\noindent(1) We design a temporal property graph data model and a query language, providing an intuitive, flexible, and effective way to model and operate PETG data. This model defines constraints in PETG data to efficiently and safely eliminate redundant information. In addition, this model is fully backward compatible with the widely used property graph data model \cite{Bonifati2018}, making it easier for users familiar with graph databases to manage PETG data.

\noindent(2) We develop a temporal graph data management system \TGraph. The system natively supports PETG data and ACID properties, equips a space-efficient temporal property storage engine that is primarily designed for Entity-History queries, and provides a fine-granularity multi-level locking mechanism to maximize its transaction throughput. 
%In addition, a declarative query language for temporal graph data is implemented to ease the system's usage.
% These designs make it easier to manage large-scale PETG data on a single machine.

\noindent(3) We test the performance of \TGraph and the existing temporal graph data management solutions on three datasets. The experimental results show that the storage space occupied by \TGraph is an average of 67\% less than the best existing solution, and its transaction throughput on HTAP workloads is 58 times higher than that of the best existing solution, while the query latency of \TGraph surpasses existing solutions by 267 times on average.
% , while the amount of query code in \TGraph is only 1/2 of that used in other data management solutions.

%The subsequent sections of this paper are organized as follows: Section \ref{sec:related-work} discusses existing temporal graph data management solutions. Section \ref{sec:model} introduces the temporal property graph model designed for property evolution. Section \ref{sec:sys-design} describes the system architecture and the design of the main functional modules. Section \ref{sec:tps-impl} illustrates the details of time implementation. Section \ref{sec:exp} reports the experimental evaluation of \TGraph, followed by conclusions. 

%Appendix \ref{sec:demo-example} shows the application case of \TGraph managing traffic temporal graph data.
\section{Related work}\label{sec:related-work}
%This section introduces existing solutions for temporal graph data management.
Based on the underlying techniques, we identify four types of solutions that are commonly used to manage temporal graph data: temporal relational database-based solutions \cite{jamkhedkar_graph_2018}, graph database-based solutions \cite{cattuto_time-varying_2013, campos_towards_2016, wu_path_2014, debrouvier_model_2021}, temporal graph databases \cite{theodorakis_aion_2024, hou_aeong_2024} and temporal graph analysis systems \cite{massri_clock-g_2022, rost_distributed_2021, byun_chronograph_2020, khurana_storing_2016, miao_immortalgraph:_2015, labouseur_g*_2015, han_chronos:_2014, cheng_kineograph:_2012}.

%\noindent\textbf{Temporal relational database}: 
\subsubsection{Temporal relational database}
The SQL:2011 standard \cite{kulkarni_temporal_2012} includes temporal features by supporting tense-related functions and a new class of tables called temporal tables, with two extra timestamp columns representing the row's time start and end.
Although the standard defines three types of temporal tables, only transaction-time temporal table is widely supported because it is easy to implement.
Nepal \cite{jamkhedkar_graph_2018} is a fault location system for VNF/SDN networks, using virtual machines in the data center as network nodes and network connections between virtual machines as edges. It uses PostgreSQL and its Temporal Tables extension for data storage. Nodes and edges are separately stored in temporal tables and are dynamically joined in the query of graph topology. For path queries, Nepal compiles them into batch SQL join queries. 
Solutions based on temporal relational databases suffer from performance degradation in topological queries due to their multi-table joins when reconstructing the graph topology.
Moreover, current solutions only support transaction time rather than valid-time, making these solutions unsuitable in scenarios where the data validity time is inconsistent with the time when the data is written to the database.
% must perform , which significantly degrades the efficiency of graph-related operations.

%\noindent\textbf{Graph database}: 
\subsubsection{Graph database}
The graph database is currently a technical solution mainly used to store and manage topology evolution temporal graph data. Cattuto \cite{cattuto_time-varying_2013} studied the method of using the Neo4j database to store and query social network information over time. It explicitly modeled time as a node and added a model transformation layer to transform temporal graph queries into graph queries. 
Debrouvier\cite{debrouvier_model_2021} models temporal properties on vertices as additional vertices attached to them, and splits edges with temporal properties to multiple edges.
Campos \cite{campos_towards_2016} stores time as a property value in the graph database, and the properties are upgraded to ``property vertices''. It also provides a SQL-like query language and a method of compiling it into a Cypher query. These solutions require extra access to multiple vertices when updating a temporal property value, which is less efficient. For queries that contain time predicates, index acceleration must be used to achieve acceptable performance. Wu \cite{wu_path_2014} provides a two-stage method to convert a temporal graph into a static graph for the temporal shortest path query. The converted graph loses some information from the original temporal graph, such as the invariance of vertices, and does not consider the property change of nodes in the temporal graph.
Graph database solutions introduce extra vertices and edges, which not only increase storage consumption but also hinder the performance of time-related operations.

\subsubsection{Temporal graph database}
Aion \cite{theodorakis_aion_2024} and AeonG \cite{hou_aeong_2024} enhance the query of temporal graphs with native graph databases by recording the history of graph updates. 
Aion \cite{theodorakis_aion_2024} highlights a critical challenge that the optimal storage strategy is highly workload- and graph-specific, leading to approaches whose performance is optimal only in parts of the query space. Aion employs two different storage structures, one for the Snapshot query and the other for the Entity-History query, resulting in redundant storage of the same data. It also introduces inefficiencies due to the overhead in the data writing process.
AeonG \cite{hou_aeong_2024} uses hybrid storage structures to manage the ``current state'' of the graph and the history of the graph separately, which introduces performance degradation for supporting temporal features.
These temporal graph databases \cite{theodorakis_aion_2024, hou_aeong_2024} primarily support transaction-time but lack support for valid-time, which is essential for representing the real-world validity of data. Additionally, these systems are primarily designed for temporal graphs characterized by topology evolution.
Moreover, solutions based on existing temporal relational, graph and temporal graph database systems are limited by the underlying system, which does not provide time-related locking mechanisms, resulting in overly coarse-grained locks and low concurrency efficiency.

%\noindent\textbf{Temporal graph analysis system}:
\subsubsection{Temporal graph analysis system}
Wang \cite{wang_time-dependent_2019} divided the temporal graph analysis system into two types of systems based on optimization goals: snapshot query and topology traversal query. Most of the earlier research was oriented to snapshot queries and aimed at rapid analysis of large-scale temporal graphs, including Kineograph \cite{cheng_kineograph:_2012}, Chronons \cite{han_chronos:_2014}, and ImmortalGraph \cite{miao_immortalgraph:_2015}. These systems consider the data locality of time and graph structure in memory and disk storage, and design iterative calculation batch scheduling algorithms based on this locality. DeltaGraph \cite{khurana_storing_2016} designs an index structure to support snapshot queries of large-scale temporal graph data. G* \cite{labouseur_g*_2015} designed a distributed index and a shared computing acceleration mechanism to support complex queries for distributed analysis and calculation of large-scale temporal graphs. ChrononGraph \cite{byun_chronograph_2020} is optimized for topology traversal queries and aims to analyze the information difference of temporal graph data at different time points. Later, new temporal graph analytical systems GRADOOP \cite{rost_distributed_2021} and Clock-G \cite{massri_clock-g_2022}  support both types of queries at the same time. These systems are all oriented towards data analysis of temporal graphs with topological evolution in which nodes or edges are continuously added or deleted, while the property values of nodes/edges almost remain unchanged, and they do not provide transaction support and fault recovery capabilities necessary for data management.

%\noindent\textbf{Summary}: 
\subsubsection{Summary}
Managing property evolution temporal graph data would involve problems of complex modeling using graph and (temporal) relational databases, poor transaction support using temporal graph analysis systems, and a lack of valid-time support using temporal graph databases. Moreover, issues like large disk space consumption and low query performance may arise because the data features and access features of PETG data are not considered in these systems. These problems raise the threshold for users to manage temporal graph data. 
% 使用现有系统管理属性演化时态图的数据也可能出现磁盘空间占用大，查询处理效率低等问题，因为现有系统并未针对性考虑属性演化时态图的数据语义及访问模式。

\section{Data model and Query Language}\label{sec:model}

This section presents the temporal property graph model and the operators and query language based on this model. We also present a set of commonly used PETG queries that can be implemented by the operators.

\subsection{Temporal Property Graph Data Model}\label{AA}
The property graph model \cite{Bonifati2018} provides a flexible and intuitive way to represent complex relationships and attributes in graph data, where both nodes and edges may have associated key-value pairs as properties. The value of properties can be of multiple types, allowing for rich and detailed descriptions. 

The temporal property graph model extends the property graph model \cite{Bonifati2018} by adding a new type of property whose value represents a temporal changed value sequence. For simplicity, we call properties whose values are not affected by time \textit{non-temporal property} and properties with temporal values \textit{temporal property}, respectively.

The timeline of a temporal property is represented by a sequence of non-decomposable, consecutive time units termed \textit{chronons} \cite{dyreson_chronon_2018}. \textit{Chronons} and its related concept \textit{time interval} \cite{goos_consensus_1998}, are derived from the domain of temporal relation database and are briefly described below.

\noindent\textbf{Definition 1 (Chronon \cite{dyreson_chronon_2018})}: A chronon $\tau$ is a nondecomposable, discrete unit of time of some fixed, minimal duration \cite{dyreson_chronon_2018}. \textit{Chronons} are of identical duration, which is defined by users when creating the temporal property. For example, a \textit{chronon} can be a second, an hour, a month, or a year. Note that a special chronon \textit{NOW} is later than any other chronon.

% Two special chronons are defined: \textit{NOW} is later than any other chronon, and \textit{INIT} is earlier than any other chronon.
%\footnote{a year may consist of 365 or 366 days, but when discussing years as non-decomposable chronons, we ignore this difference and regard them as ``identical''.}

\newcommand{\tiau}{\ensuremath{\overline\tau}}
\noindent\textbf{Definition 2 (Time Interval \cite{goos_consensus_1998})}: A time interval $\tiau=[\tau_s, \tau_e)$ on a timeline is formed by a series of continuous chronons \cite{goos_consensus_1998}, representing the duration of time starting from chronons $\tau_s$ (inclusive) to $\tau_e$ (exclusive), satisfying $\tau_s <\tau_e$.
%Note that a time interval $[\tau_s, \tau_e)$ equals a chronon $\tau_s$ if $\tau_e$ is the next chronon after $\tau$ along the timeline.

\begin{table*}[tb]
\centering
\footnotesize
\caption{Operators of temporal property graph}
\label{tab:access-interface}
\begin{tabular}{@{}lll@{}}
\toprule
Operation Target & Operators Definition & Description \\
\midrule
\multirow{8}*{Topology} & $newNode(l_0, l_1, \cdots)\Rightarrow v$ & create a vertex $v$ with labels $l_0, l_1, \cdots$ marking it\\
& $newRel(v_s, v_d, l_0, l_1, \cdots)\Rightarrow r$ & create an edge $r$ from $v_s$ to $v_d$ with labels \\
& $getNode(l_0, l_1, \cdots)\Rightarrow [v, \cdots]$  & return all vertics whose labels matching $l_0, l_1, \cdots$\\
& $getRel(v, l_0, l_1, \cdots)\Rightarrow [r, \cdots]$  & return all edges of $v$ whose labels matching $l_0, l_1, \cdots$\\
& $getStart(r)\Rightarrow v_s$              & return $r$ start node $v_s$                   \\
& $getEnd(r)\Rightarrow v_d$                & return $r$ end node $v_d$                    \\
& $delNode(v)$                            & delete $v$ and its connected edges      \\
& $delRel(v_s, v_d, l_0, l_1, \cdots)$   & delete all edges from $v_s$ to $v_d$ whose labels matching $l_0, l_1, \cdots$ \\
\hline
\multirow{3}*{Non-temporal Property} & $setProp(v|r, p, val)$   & set $v$ (or $r$)'s non-temporal property $p$'s value to $val$ \\
& $getProp(v|r, p)\Rightarrow val$    & return $v$ (or $r$)'s non-temporal property $p$'s value  \\
& $rmProp(v|r, p)$    & delete the non-temporal property $p$ on $v$ (or $r$)  \\
\hline
\multirow{3}*{Temporal Property} & $setTP(v|r, tp, \tiau, val)$  & set $v$ (or $r$)'s temporal property $tp$'s value to $val$ during $\tiau$ \\
& $getTP(v|r, tp, \tiau)\Rightarrow \Psi$ & return $v$ (or $r$)'s temporal property $tp$'s Time Interval Series $\Psi$ during $\tiau$\\
& $rmProp(v|r, tp)$    & delete the temporal property $tp$ on $v$ (or $r$)  \\
\hline
\multirow{5}*{Time Interval Series} & $slice(\Psi, \tiau)\Rightarrow \Psi_{sub}$ & return a truncate Time Interval Series whose time within $\tiau$\\
& $valueAt(\Psi, \tau)\Rightarrow value$ & return the value of $\Psi$ at time $\tau$\\
& $valueList(\Psi, \tiau)\Rightarrow [value, \cdots]$ & return the value list of $\Psi$ during $\tiau$\\
& $set(\Psi, \tiau, val)\Rightarrow \Psi_{new}$ & change $\Psi$ value to $val$ during $\tiau$\\
& $entry(\Psi)\Rightarrow [(\tiau_{slice}, val), \cdots]$ & access $\Psi$ as a list of $(\tiau_{slice}, val)$ entries\\
\hline
\multirow{4}*{Time Interval} & $intersection(\tiau_1, \tiau_2, \cdots, \tiau_n)\Rightarrow \tiau_{new}$ & returns the intersection of two or more time intervals (brief as $\cap$)\\
& $union(\tiau_1, \tiau_2, \cdots, \tiau_n)\Rightarrow \tiau_{new}$ & returns a time interval of multiple continuous time intervals (brief as $\cup$)\\
& $diff(\tiau_1, \tiau_2)\Rightarrow \tiau_{new}$ & return the difference of two time intervals\\
& $shift(\tiau_{old}, \tau)\Rightarrow \tiau_{new}$ & move the start point of $\tiau_{old}$ to $\tau$ without duration change\\
\bottomrule
\end{tabular}
\end{table*}

\noindent\textbf{Definition 3 (Time Interval Series)}: A \textit{Time Interval Series} $\Psi$ is an ordered set composed of $(\tiau, value)$ tuples: 
$$\Psi= \langle(\tiau_1, value_1), (\tiau_2, value_2), \cdots, (\tiau_n, value_n )\rangle$$

The tuple $(\tiau_j, value_j)$ indicates that the value in the time interval $\tiau_j=[\tau_{j_s}, \tau_{j_e})$ is $value_j$, for $1\leq j\leq n$. $\tiau_1, \cdots, \tiau_n$ are ascending time intervals without overlap, satisfying $\tau_{j_s}< \tau_{j_e}\leq\tau_{{j+1}_s}$, for $1\leq j<n$. 
Note that $\tau_{n_e}$ can be $Now$, which means that the value at any time after $\tau_{n_s}$ is $value_n$.
% $t_{s_1}$ can be $Init$, which means that its value is $value_1$ at any time before $t_{e_1}$.
%The \textit{Time Interval Series} utilizes ``Point-based'' semantics \cite{jensen_temporal_2009}, which means that the time intervals in a tuple signify the validity of the corresponding values, regardless of the original validity periods entered by the user. If there is a situation where the time intervals of two adjacent tuples is continuous, and their values are identical, say $\Psi_1 = \langle ([\tau_a, \tau_b), value), ([\tau_c, \tau_d), value)\rangle$ and $\tau_b=\tau_c$, then $\Psi_1$ are considered to be semantically equivalent to a coalesced \textit{Time Interval Series} $\Psi_2 = \langle([\tau_a, \tau_d), value)\rangle = \Psi_1$.
For example, the ``status'' temporal property of the road in Fig. \ref{fig:traffic-model} can be represented by a \textit{Time Interval Series} $\Psi = \langle([``\text{8:00}", ``\text{8:15}"), ``slow"), ([``\text{8:15}", ``\text{8:20}"), ``jam")$, $([``\text{8:20}", ``\text{8:45}"), ``smooth"), ([``\text{8:45}", Now), ``slow")\rangle$.

The \textit{Time Interval Series} extends the commonly used concepts of ``Time Series \cite{liu_time_2009-1}'' and ``Time Sequences \cite{liu_time_2009}'' that are typically defined as a set of tuples $\langle v, t\rangle$, where $v$ represents a data object and $t$ represents a time point signifying when the data object changes \cite{liu_time_2009-1}.
% The \textit{Time Interval Series} replaces the time point $\tau$ with a time interval $\tiau$ in each tuple. 
By replacing the time point $\tau$ with a time interval $\tiau$ in each tuple, users can directly define the value of a time interval $\tiau$ rather than specifying the value on every chronon $\tau$ satisfying $\tiau_s\leq\tau<\tiau_e$.
%关于Time Interval Series扩展了Time Series和Time Sequences这段有3个问题：1、为什么要专门说这件事；2、我没有看懂TIS是怎么扩展TS的（也可能是我的问题）；3、TS中的t是时间点，需要在表述中显式地转换为chronon。（i.e. Chronon）

The \textit{Time Interval Series} utilizes the ``Point-based semantics'' \cite{jensen_temporal_2009, point_based_2004}, meaning the data can be interpreted as a sequence of states indexed by points in time (i.e., Chronons). This indicates that the decomposition of time intervals into chronons does not affect the semantics of data \cite{point_based_2004} (and vice versa). For example, $\langle([1, 4), v)\rangle$ are semantically equal to $\langle(1,v)$, $(2,v)$, $(3,v)\rangle$ (if the duration of a chronon is 1). If the time intervals of two adjacent tuples are continuous and their values are identical, say $\Psi_1 = \langle ([\tau_a, \tau_b), v)$, $([\tau_c, \tau_d), v)\rangle$ and $\tau_b=\tau_c$, then $\Psi_1$ are considered semantically equivalent to a coalesced \textit{Time Interval Series} $\Psi_2 = \langle([\tau_a, \tau_d), v)\rangle = \Psi_1$.
Based on this semantics, redundant tuples in the \textit{Time Interval Series} can be coalesced naturally to save space, particularly when value changes in PETG data occur less frequently than event updates. This scenario is common when updates in PETG data are automatically generated by IoT sensors at regular intervals.

\noindent\textbf{Definition 4 (Temporal property graph)}: A temporal property graph is defined as $TG=(V,E)$. $V$ is a set of vertices $v=(vid, L, P, TP)$, where $vid$ is the identifier of vertex $v$, which uniquely determines $v$ in $V$; $L$ is the set of label $l$ marking $v$; $P$ is the set of (non-temporal) property $p=(name, value)$ of $v$ where $name$ is the name of the property and $value$ is the property value; $TP$ is the set of temporal properties $tp = (name, \Psi)$ of $v$ where $\Psi$ is a \textit{Time Interval Series}. $E$ is the set of edges $e=( eid, v_s, v_e, L, P, TP)$, where $eid$ is the identifier of $e$; $v_s$ is the starting vertex of $e$; $v_e$ is the end vertex of $e$; $L$ is the label set of $e$, $P$ and $TP$ are the non-temporal and temporal property sets of $e$ respectively. Multiple edges are allowed between two vertices. 

Notes that:
\noindent(1) Topological changes in a temporal graph can also be modeled by adding a temporal property to the vertices/edges representing their valid time.
\noindent(2) Different temporal properties can have chronons with different durations. For example, a ``travel-time'' temporal property of an edge (representing a road) may use chronons whose duration is a minute. But a ``passable'' temporal property on the edge may use chronons whose duration is an hour.

%%%%%%%%%%%%%%%%%%%%%%%%%%%%%%%%%%%%%%%%%%%%%%%%%%%%%%%%%%%%%%%%%%%%%%%%%%
%\begin{figure}[tb!]
%\centerline{\includegraphics[width=\columnwidth]{figures/tp-model.pdf}}
%\caption{Temporal Property Graph Model.}
%\label{fig:sys-data-model}
%\end{figure}

%\noindent\textbf{Example 2}: Fig. \ref{fig:sys-data-model} illustrates an example temporal property graph with 5 nodes ($A$ to $E$), 8 edges, and 3 temporal properties ($tp_1$ to $tp_3$). $tp_1$ and $tp_2$ are temporal property of edge $x$, $tp_3$ is the temporal property of node $C$. The right side of the picture shows the temporal dimension of the three temporal properties, whose values change according to time.
%%%%%%%%%%%%%%%%%%%%%%%%%%%%%%%%%%%%%%%%%%%%%%%%%%%%%%%%%%%%%%%%%%%%%%%%%%

\subsection{Operators of Temporal Property Graph}\label{db-interface}
We define a set of basic operators for the temporal property graph model in Table \ref{tab:access-interface}. Based on their operational targets, operators are categorized into five groups: (1) Topology, (2) Non-Temporal Property, (3) Temporal Property, (4) \textit{Time Interval Series}, and (5) Time Interval.
The operators in the ``Topology'' and ``Non-temporal property'' categories are the same as the operators in the property graph model \cite{nadime_francis_cypher_2018}, providing backward compatibility. Operators in the other categories are newly added to handle reads and writes of temporal properties and related \textit{Time Interval Series} and time intervals. 

We next briefly introduce two fundamental operators, \i.e., $setTP()$ and $getTP()$.
Operator $setTP()$ allows updating the \textit{Time Interval Series} of a temporal property and guarantees the ``Point-based Semantics'' \cite{point_based_2004} of the \textit{Time Interval Series}. 
Operator $getTP()$ retrieves a \textit{Time Interval Series} slice from the specified temporal property. The slice of \textit{Time Interval Series} can be further accessed as an ordered list of $(\tiau, v)$ triples by $entry()$.  So in the example of traffic temporal graph data (Fig. \ref{fig:traffic-model}), after invoking $setTP(road, ``status", ``\text{08:18}"$, $Now, ``jam")$, the call of operator $getTP(road, ``status"$, $``\text{08:10}", Now)$ would return a \textit{Time Interval Series} $\Psi = \langle([``\text{08:10}",$ $``\text{08:15}"), ``slow"), ([``\text{08:15}", Now), ``jam")\rangle$.

\subsection{Queries of Temporal Property Graph }
\label{sec-query}
Using the temporal property graph operators, we can implement more complex operations on PETG data. 
The following are some common temporal graph operations (also used in our experimental study), building on these operators:

\noindent(1) \textit{Update} sets the value of $tp$ on entity $v|r$ during $\tiau$ to $v$, i.e., a single $setTP()$ operation. 
%Statement (i) in Table \ref{tab:demo-tcypher} sets the travel time at the specified time interval to 36s for ``Haidian East Road''.

\noindent(2) \textit{Append} is a sequence of $setTP()$ operations on $tp$ of an entity $v|r$, i.e., $[setTP(v|r, tp, \tiau_1, val_1)$, $\cdots$, $setTP(v|r, tp, \tiau_n, val_n)]$, satisfying $\tiau_i\cap\tiau_{i+1}=\varnothing$ and $\tau_{i_s}<\tau_{{i+1}_s}$ for $1\leq i<n$. \label{append_intro}
Additionally, if $\Psi$ of the property on the entity is not an empty set, \textit{Append} should also satisfy $\tau_{max} < \tau_{1_{s}}$ where $\tau_{max}$ is the latest non \textit{NOW} chronon in $\Psi$.

\noindent(3) \textit{Entity-History} queries the history of $tp$ on an entity $v|r$ during $\tiau$, i.e., a single $getTP()$ operation. 
%Statement (ii) in Table \ref{tab:demo-tcypher} queries the history of jam status for ``Haidian East Road'' during the specified time range. 

\noindent(4) \textit{Snapshot} queries the values of $tp$ on all entities in a graph at a specified chronon $\tau$, implemented by iterating all entities and calling the $getTP(e, tp, \tau)$ on each entity. 
%Statement (iii) in Table \ref{tab:demo-tcypher} queries the jam status of all roads at the specified time.

%\noindent(5) \textit{Aggregate Temporal Property (ATP)} is a special Entity-History query that performs aggregation $f(\Psi)$ on the result \textit{time interval series} $\Psi$ of the Entity-History query $getTp(v|r, tp, \tiau)$. The aggregate function $f$ can be ``min'', ``max'', ``count'' or user-defined aggregation functions. 
\noindent(5) \textit{Graph Aggregate Temporal Property (GATP)} performs aggregation $f(\Psi)$ on all entities in the graph, where \textit{time interval series} $\Psi$ is the result of query $getTp(v|r, tp, \tiau)$ on each entity of $tp$ during $\tiau$. The aggregate function $f$ can be ``min'', ``max'', ``avg'' or user-defined aggregation functions. 
%Statement (iv) in Table \ref{tab:demo-tcypher} queries the maximum travel time of all roads during the specified time range. 

\noindent(6) \textit{Entity Temporal Property Condition (ETPC)} queries entities (nodes or edges) whose value of temporal property $tp$ is within the specified value ranges $[v_{min}, v_{max}]$ during specified time interval $\tiau$. 
%Statement (v) in Table \ref{tab:demo-tcypher} queries all roads whose travel time is between 600s and 1200s during the specified time range. 
% Statement (6) creates an index on the ``travel\_time'' property, speeding up the ``value within'' queries like the one in statement (5).

\begin{algorithm}[tb]
\footnotesize
\caption{$ReachableArea(TG, v_s, \tau_s, dur)$}
\label{algo:reachable}
\DontPrintSemicolon
% \KwIn{Temporal Property Graph $TG=(V,E)$, source node $s \in V$, departure time $t_s$, and timeout duration $dur$}
\KwOut{$\Gamma$ (earliest arrival time of nodes that are reachable from $v_s$ within $dur$ departing at $\tau_s$).}
\vspace{0ex}
% \KwResult{$\Gamma$}
% \SetKwData{Left}{left}\SetKwData{This}{this}\SetKwData{Up}{up}
\SetKwFunction{SCN}{earliestArriveNode}
\SetKwFunction{UNA}{earliestArrTime}
\SetKwProg{Fn}{Function}{:}{}
\BlankLine                                    % 加一行空行
$\Omega\leftarrow \varnothing$; % \tcp*[f]{\small Nodes in Calculating status}\;
$\Delta\leftarrow \varnothing$; % \tcp*[f]{\small Nodes in Calculated status}\;
$\Gamma\leftarrow \varnothing$; % \tcp*[f]{\small Nodes' earliest arrive time}\;
\tcp*[f]{\small Init}\;
$\Omega \leftarrow \Omega\cup v_s$;
$\Gamma[v_s] \leftarrow \tau_s$;
$\tau_{max} \leftarrow \tau_s+dur$\;

\While{$\Omega\neq\varnothing$}{
    $v_n \leftarrow $ \SCN{$\Omega, \Gamma$}\;
    $\Omega \leftarrow \Omega - v_n$;
    $\Delta \leftarrow \Delta\cup v_n$\;
    \If{$\Gamma[v_n] < \tau_{max}$}{
        \ForEach{$r\leftarrow getRel(v_n, outgoing)$}{
            $v\leftarrow getEnd(r)$\;
            \If{$v \in \Omega$}{
                $\tau_v\leftarrow$ \UNA{$r, \Gamma[v_n], \tau_{max}$}\;
                \lIf{$\tau_v < \Gamma[v]$}{ $\Gamma[v]\leftarrow \tau_v$ }
            }\ElseIf{$v \notin \Delta$}{
                $\tau_v\leftarrow$ \UNA{$r, \Gamma[v_n], \tau_{max}$}\;
                $\Omega \leftarrow \Omega\cup v$;
                $\Gamma[v] \leftarrow \tau_v$
            }
        }
    }
}
% \KwRet $\Gamma$\;
\Fn{\UNA{$r, \tau_0, \tau_1$}}{
    $arriveT\leftarrow \tau_1$\;
    $\Psi\leftarrow getTP(r, $ ``travel time"$, [\tau_0, \tau_1))$\;
    \ForEach{$([\tau_a, \tau_b), val)\leftarrow entry(\Psi)$}{
        $arriveT\leftarrow min(arriveT, \tau_a+val)$
    }
    \KwRet $arriveT$\;
}
\vspace{-1ex}
\end{algorithm}

\noindent(7) \textit{Reachable Area} returns all nodes $v\in V$ (and their earliest arrival time $\Gamma[v]$) that can be reached ($v\in\Gamma$) within a given time duration $dur$, departing from a specific node $v_s$ at a given time $\tau_s$. 
Algorithm \ref{algo:reachable} shows an implementation of the calculation process of a reachable area query using the temporal property graph model and its operators. The process is delivered from the earliest arrival path algorithm for the temporal road graph \cite{dreyfus_appraisal_1969}. 
The earliest arrival time of road $r$ is calculated by querying its ``travel time'', which is a function of departure time $\tau_0$. The function is packed by a \textit{Time Interval Series} $\Psi$ fetched by $getTp()$.
The \texttt{earliestArriveNode()} returns a node $v_n\in\Omega$ whose arrival time is earliest according to $\Gamma$ (not defined in the code because it does not contain operators of the temporal property graph).

\subsection{Temporal Query Language}
We develop a high-level query language, named TCypher, to support temporal property graph queries. 
TCypher is extended from \textit{Cypher} \cite{nadime_francis_cypher_2018}, a popular declarative query language of graph databases.
%, aimed at providing efficient access to and manipulation of graph-structured data through its intuitive syntax. 
% The language was originally developed by the Neo4j community and has now become an open standard for graph database interaction. 
%TCypher is a database query language supporting the temporal property graph model (Section \ref{sec:model}) that extends and is compatible with the Cypher language. 
We try to minimize the modifications so users familiar with Cypher can easily use TCypher to manipulate temporal graphs.
Table \ref{tab:demo-tcypher} are some example statements of TCypher with respect to the queries of Section \ref{sec-query}. 

\begin{table}[tb!]
\centering
%\scriptsize
\footnotesize
\caption{TCypher example queries managing a traffic graph}
\label{tab:demo-tcypher}
\vspace{-1ex}
\begin{tabular}{p{8.2cm}}
\toprule
\begin{lstlisting}
#Statement (i) Update a temporal property
MATCH ()-[road {name: "Haidian East Road"}]->() SET road.travel_time = TIS("2010-05-01 08:00"~"2010-05-01 08:10": 36)
\end{lstlisting}\\
% \hline
% \begin{lstlisting}
% #Statement (ii) Entity History Query
% MATCH ()-[road {name: "Haidian East Road"}]->() RETURN TIS_SLICE(road.jam_status, "2010-05-01 08:00"~"2010-05-01 08:30")
% \end{lstlisting}\\
% \hline
% \begin{lstlisting}
% #Statement (iii) Snapshot Query
% MATCH ()-[road]->() RETURN road.name, TIS_AT(road.jam_status, "2010-05-01 08:00")
% \end{lstlisting}\\
\hline
\begin{lstlisting}
#Statement (ii) GATP(MAX) Query
MATCH ()-[road]->() RETURN road.name, TIS_AGGR_MAX(road.travel_time, "2010-05-01 08:00"~"2010-05-01 08:30")
\end{lstlisting}\\
\hline
\begin{lstlisting}
#Statement (iii) ETPC Query
MATCH ()-[road]->() WHERE TIS_WITHIN(road.travel_time, '2010-05-01 08:00'~'2010-05-01 08:30', 600, 1200) RETURN road.name
\end{lstlisting}\\
\bottomrule
\end{tabular}
\end{table}

TCypher mainly expands upon Cypher in two ways: (1) It introduces two new basic data types for describing Time Intervals and \textit{Time Interval Series}. For example, \textit{Time Interval Series} can be used as property values when defining properties (\texttt{TIS} expression in statement (i), Table \ref{tab:demo-tcypher}). (2) It incorporates many functions that manipulate \textit{Time Interval Series} and construct predicates (\texttt{TIS\_WITHIN} expression in statement (iii), Table \ref{tab:demo-tcypher}) for querying temporal properties in a temporal graph. Note that TCypher also supports user-defined functions (UDF), allowing complex queries to be implemented as functions that can be called in the TCypher language.

% \noindent\textbf{Example 2}: Table \ref{tab:demo-tcypher} provides example statements showing how to modify, query temporal graph data, and create indexes using the TCypher language. Statement (1) sets the travel time at a specified time interval to 36s for ``Haidian East Road''. Statement (2) queries the history of jam status for ``Haidian East Road'' during a specified time range (Entity-History query). Statement (3) queries the jam status of all roads at a specified time (Snapshot query). Statement (4) queries the maximum travel time of all roads during a specified time range. Statement (5) queries all roads whose travel time is between 600s and 1200s during the specified time range. Statement (6) creates an index on the ``travel\_time'' property, speeding up the ``value within'' queries like the one in statement (5).

\section{System design}\label{sec:sys-design}
This section presents the design of \TGraph.
%, an effective and efficient single-machine data management system for temporal graph data.
%, a temporal graph data management system,

\subsection{Design Goals and Principles}
\TGraph is a single-machine PETG data management system with the following design goals and principles.

\noindent\textit{(1) Native support for valid-time temporal property graph model}. \TGraph natively supports the temporal property graph model, providing an intuitive, flexible, and effective way to operate PETG data. The valid time of data is specified by users, allowing for modification and analysis of existing data.

\noindent\textit{(2) ACID properties and high HTAP transaction throughput}. \TGraph should support user-level transactions. %Improve  transaction throughput by reducing the occurrence of deadlocks and mutual waiting caused by the workload pattern of PETG data.
Moreover, \TGraph should support a time-related concurrency control mechanism to improve transaction throughput.

\noindent\textit{(3) Compatibility with current graph databases}. \TGraph should provide (API and file format) compatibility with Neo4j, the most popular graph data management system.
This implies \TGraph is also capable of managing the (non-temporal) graph databases created by Neo4j. When the recording of graph dynamics becomes a user requirement, the DBMS running on the database can be transitioned from Neo4j to \TGraph seamlessly. This compatibility plays a crucial role in reducing user expenses.
% For this reason, we have selected the popular graph database Neo4j as the foundational graph database system for building \TGraph and extensively extended and modified it. 

%\noindent\textit{(4) Space/time efficient storage for append write and entity history queries}. To manage large-scale property evolution temporal graph data on a single machine, \TGraph employs several techniques that reduce the redundancy of data and support efficient Entity-history queries of temporal graph data.
\noindent\textit{(4) Space/time efficient storage for frequent Append write and Entity-History query}. \TGraph should employ techniques to reduce data redundancy and support efficient \textit{Entity-History} and \textit{Append} write, which are the most frequent read/write queries of PETG data.

%\noindent\textit{(5) Easy of use interface and query optimal with index}. A high-level declarative query language TCypher (based on Neo4j's Cypher language) is designed, and its compiler is implemented to process high-level temporal graph queries. 
\noindent\textit{(5) Easy of use}. \TGraph should support a high-level declarative query language to facilitate users in developing complex temporal graph queries.

\subsection{Architecture}
\begin{figure}[tb!]
\centerline{\includegraphics[width=0.84\columnwidth]{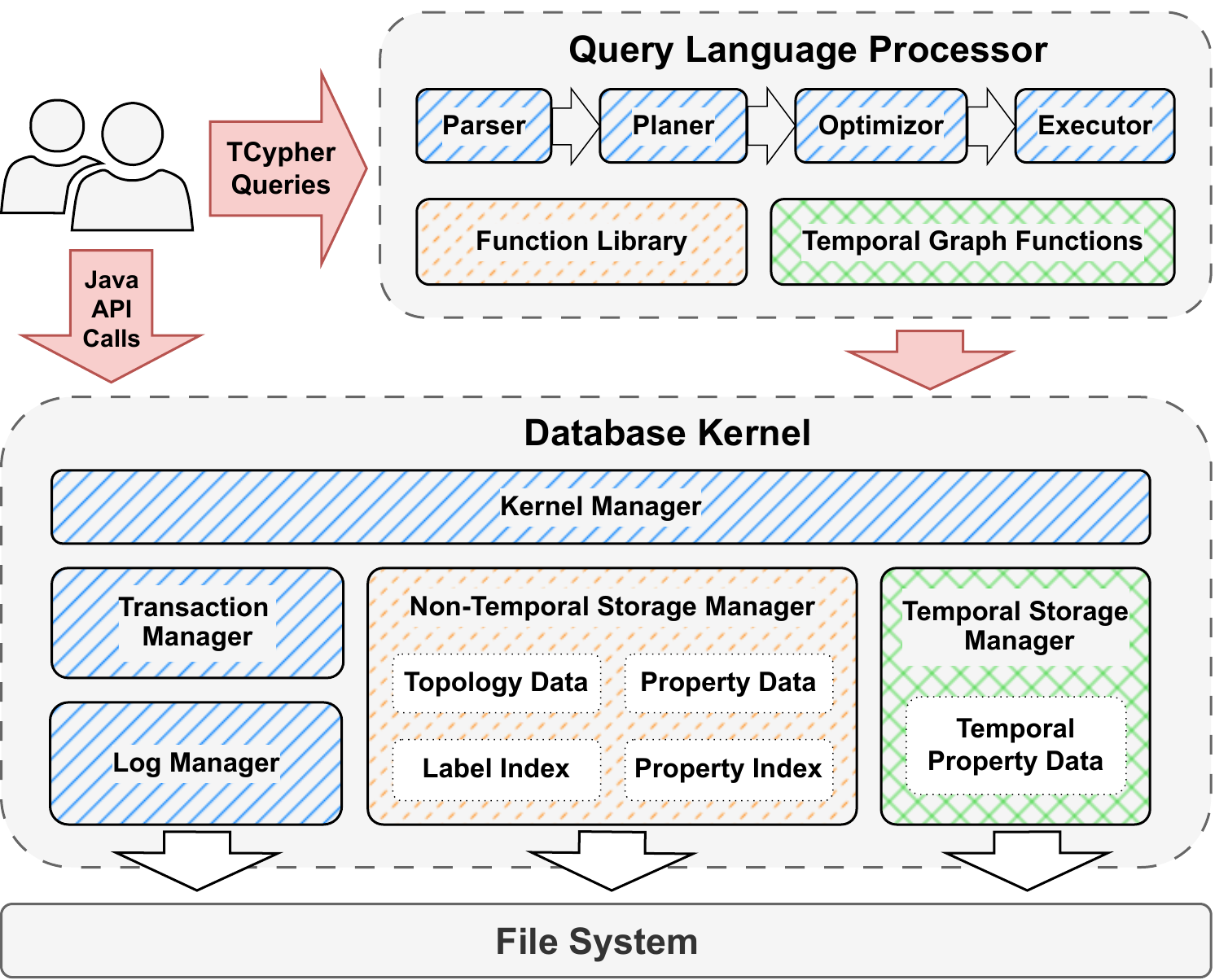}}
\vspace{-1ex}
\caption{System architecture of \TGraph}
\vspace{-2ex}
\label{fig:sys-arch}
\end{figure}

%The architecture of \TGraph, derived by extending and modifying the architecture of Neo4j, is illustrated in Fig. \ref{fig:sys-arch}. 
\TGraph extends the architecture of Neo4j to maintain compatibility with current graph databases. The architecture of \TGraph is illustrated in Fig. \ref{fig:sys-arch}.
%Despite adding new modules (boxes filled with green cross-hatches) to manage temporal property data, a number of Neo4j database modules (boxes filled with blue hatches) are refactored to involve the capacity of temporal property access.
%Some other modules (boxes filled with orange dashed lines) remain unchanged. 
Dedicated temporal property modules (indicated by green cross-hatches) are introduced, while several legacy Neo4j modules (shown with blue hatches) are refactored to incorporate valid-time temporal graph access capabilities. Remaining modules (marked with orange dashed lines) are kept unchanged.

\TGraph consists of two main components: the \textit{database kernel} and the \textit{query language processor}. The {database kernel} provides comprehensive temporal graph data management and transactional capabilities that are operable through Java APIs. The {query language processor} is an optional component. When loaded, the system can handle TCypher queries, compiling them into Java procedures and calling the database kernel for execution. 

%\noindent\textbf{Database kernel} 
\subsubsection{Database kernel}
It implements the data access interface defined in Section \ref{db-interface} and other transaction management interfaces as Java APIs for users to operate the database. It contains five modules: (1) the {kernel manager} acts as a query portal and manages important control flows in the database; (2) the {non-temporal storage manager} manages graph topology, non-temporal property data and its indexes; (3) the {temporal storage manager} manages temporal property data; (4) the {transaction manager} provides a locking mechanism and private memory buffers for implementing the ACID properties of the database; and (5) the {log manager} controls transaction logging and fault recovery abilities.

The kernel manager is the core part of the database kernel. It uniformly manages all requests that read, write, and manage the database instance, oversees all internal states of the database instance, and controls all processes across modules. It invokes functionalities from other modules to ensure the ACID properties of database access.

The storage manager manages graph topology data, non-temporal property data, and temporal property data. The temporal and non-temporal properties share the same namespace. The temporal property data module, primary storage of temporal property data, supports efficient \textit{Entity-History} query and \textit{Append} write, which are the most frequent read/write queries of PETG data. The module also minimizes the space cost by eliminating data redundancy.

The transaction manager and the log manager work together to ensure the system's ACID properties and support log-based fault recovery. We also extend the transaction manager to support more efficient concurrent queries of temporal graphs by refactoring its locking mechanism, which allows fine-granularity locks to be added at specified time intervals.

%\noindent\textbf{Query language processor} 
\subsubsection{Query language processor}
It is an optional module of \TGraph that generates query plans of TCypher queries and compiles them into Java procedures that internally invoke the database kernel and then execute the procedures. To this end, we extend the query parser and function library of the original Cypher language to support TCypher's syntax and semantics, and implement temporal-property-related query operators in the execution engine.

\subsection{Temporal Property Storage}\label{sec:tps}
%Compared to other temporal graph data, 
% The most common query related to temporal properties in PETG data is the Entity-History query, which To efficiently support this query, we have designed the Temporal Property Storage module for \TGraph.

The primary design goals of the temporal property data storage module of \TGraph are: (1) efficient handling of the \textit{Entity-History} query (see Section \ref{sec-query}) that examines the historical changes of a known property on a certain node/edge over a specified period of time; (2) accommodating fast data writes, especially the \textit{Append} writing of temporal properties for each node/edge, and (3) reducing disk space usage. 
In the following discussion, we assume the data items of temporal property storage are in the form of $\langle e, tp, \tiau, v\rangle$, where $e$ is a vertex/edge, $tp$ is a temporal property, $\tiau=[\tau_s, \tau_e)$ represents the valid time interval of the data item, and $v$ is the corresponding value during $\tiau$.
 % and minimizing write amplification
% In addition, the temporal property storage must support the \textit{Schema Free} feature of Neo4j, allowing the creation of new properties without modifying existing data and allowing different nodes/edges to have different temporal properties. This provides flexibility for modeling in data management.

There has been works \cite{time_index_1990,period_index_2019} on access structures of time interval data, but they are less efficient for the Entity-History query, as their query predicates do not contain the entity as a query parameter, with which the non-temporal access structures (such as B-Tree \cite{btree_survey_1979} or LSM-Tree \cite{oneil_log-structured_1996}) can be used to efficiently answer the Entity-History query by ordering data by $\langle e, tp, \tau_s\rangle$. B-Tree \cite{btree_survey_1979} and LSM-Tree \cite{oneil_log-structured_1996} provide $O(log(N))$ complexity when querying from $N$ data items. Moreover, LSM-Tree \cite{oneil_log-structured_1996}  utilizes the sequential write features of disks for high writing throughput and is capable of handling write-intensive workloads like \textit{Append} in PETG.

However, data skew in the PETG data on the property and time dimension is not considered in LSM-Tree \cite{oneil_log-structured_1996}. 
Properties with low update frequency have fewer records and are sparse in all temporal property data, resulting in inefficient access and difficulties in compressing the data. In addition, the frequently updated recent data can have a smaller file size (or more levels), trading off read efficiency for write performance.
% Querying the temporal property data that is far from the current time, with few updates, can be accelerated by limiting the levels of the LSM-Tree.
% Data farther from the current time exhibits lower update frequency, yet each LSM-Tree compaction repeatedly reads it into memory and writes it back to disk, incurring substantial unnecessary computation and I/O waste. Queries on such data must first access the newly merged files before reaching the old files that actually store the data, while all accesses to the new files constitute pure overhead. 
Therefore, we designed a tree-based data structure called ``Temporal Interval Merge Tree (TIM-Tree)'' that drew on the idea of LSM-Tree \cite{oneil_log-structured_1996} but introduced extra partition mechanisms in the property and time dimension.
% Those goals and constraints require that temporal property data be organized as time-ordered, that disk data be written in batch mode as much as possible, and that data redundancy be identified and eliminated. Based on these requirements as well as data access characteristics, we drew on the idea of LSM-Tree \cite{oneil_log-structured_1996} and KD-Tree \cite{robinson_k-d-b-tree:_1981} and designed a tree-based data structure called ``Temporal Interval Merge Tree (TIM-Tree)'' oriented for \textit{Entity-History} queries on the \textit{Time Interval Series} data. The TIM-Tree organizes data hierarchically, with different levels of the tree indexing different dimensions of the data items.

% In SCSM-Tree, adjacent data items with the same values (states) might be compacted to save space. The architecture (shown in Fig. \ref{fig:tp-arch}) is composed of memory and disk parts.

%\noindent\textbf{TIM-Tree Data Structure}:
\subsubsection{TIM-Tree Data Structure}
\begin{figure}[tb!]
\centerline{\includegraphics[width=0.82\columnwidth]{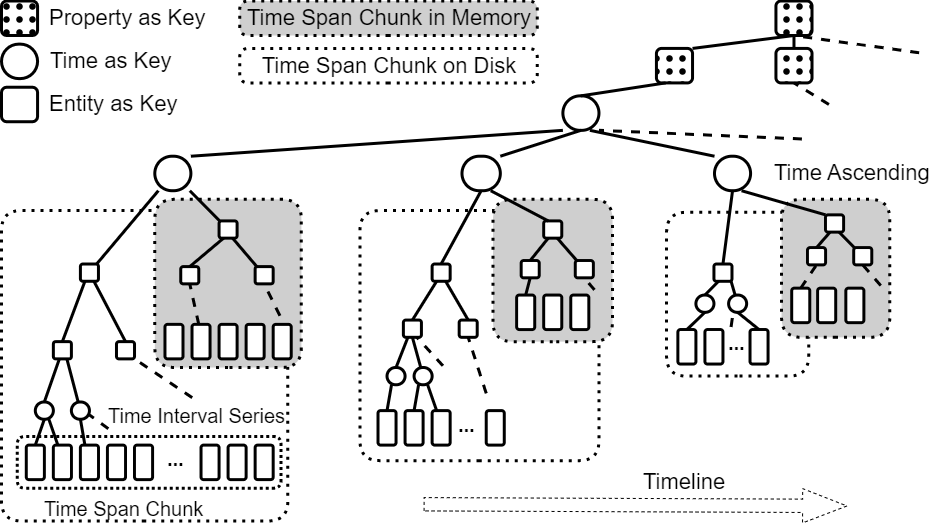}}
\vspace{-1ex}
\caption{The structure of Time Interval Merge Tree}
\vspace{-2ex}
\label{fig:tps-struct-example}
\end{figure}

Fig. \ref{fig:tps-struct-example} shows the data structure of TIM-Tree.
% The leaf nodes of TIM-Tree are data items of \textit{Time Interval Series} (rectangles in Fig. \ref{fig:tps-struct-example}).
% , which are in the form of $\langle entity$, $property$, $timeStart$, $timeEnd$, $value\rangle$. 
The data items of the TIM-Tree are first partitioned/ordered by $tp$ (square nodes with dots) and then by $\tiau$ (circle nodes). This method efficiently narrows the search space based on the specified properties and time intervals. Data items inside the partition are then grouped as \textit{Time Span Chunks} (dashed-line boxes). \textit{Time Span Chunks} of a temporal property divide the data into a series of across a series of disjoint, ordered, and contiguous time ranges along the timeline.
The chunk size decreases as time increases (a tunable parameter specified by the user) in most cases, which tends to store the most recent data, typically updated more frequently, in smaller chunks. The nodes in the upper levels of the TIM-Tree, which index \textit{Time Span Chunks} by $tp$ and $\tiau$, are stored in memory, reducing the overhead of disk I/O operations during the initial search phase and allowing for quick filtering based on these primary dimensions.
Inside the \textit{Time Span Chunks}, data items are of a single property and a range of time; they are ordered by $\langle e, \tau_s\rangle$ in ascending order and can be interpreted as an ordered set of \textit{Time Interval Series} (rectangles grouping leaf nodes) for each $e$.
% , and stored as an LSM-Tree \cite{oneil_log-structured_1996}. 
A \textit{Time Span Chunk} can be stored either on disk or in memory (dashed-line boxes with gray background); the latter corresponds to their counterparts on disk (sibling dashed-line boxes with white background) and are used in updating the TIM-Tree.
Each in-memory \textit{Time Span Chunk} and its counterpart chunk on disk shares the same property and time range; they correspond to the in-memory component and disk component of the LSM-Tree \cite{oneil_log-structured_1996}.
% of two  (one), within which data items are ordered by $\langle e, \tau_s\rangle$. Note that: (1) Every temporal property has a special \textit{Time Span Chunk} $LC$ (last chunk) whose start chronon equals the end chronon of the property’s last regular chunk (or 0 if none) and whose end chronon is NOW; $LC$’s on-disk component is always empty; (2) The in-memory component of any \textit{Time Span Chunk}, including $LC$, may also be empty.
% Like KD-Tree \cite{robinson_k-d-b-tree:_1981}, different levels index different dimensions of the records. The indexing dimensions of the TIM-Tree from the top down are (1) Property, (2) Time Interval, (3) Entity, and (4) Time Interval. 
% The nodes in the upper levels of the TIM-Tree are indexed first by property  and then by time . Moreover, these nodes are stored in memory, reducing the overhead of disk I/O operations during the initial search phase and allowing for quick filtering based on these primary dimensions.
% The intermediate and leaving nodes in the low levels of the tree, grouped as ``Time Span Chunks'' (dashed line), are first indexed by entity (square nodes) and then by time (circle nodes). These chunks can be stored either on disk or in memory. Each \textit{time span chunk} covers data over a specific time range, with the chunk size decreasing as time increases. This ensures that the most recent data, which is typically updated more frequently, is stored in smaller chunks.

\subsubsection{Data Items coalescing and splitting}
Based on the ``Point-based semantics" \cite{point_based_2004} of the \textit{Time Interval Series}, multiple data items $\langle e, tp, \tiau_0, v\rangle\cdots\langle e, tp, \tiau_k, v\rangle$ can be merged (coalesced) into a single data item $\langle e, tp, \tiau, v\rangle$ if they are in the same \textit{Time Span Chunk} and have continuous $\tiau_i$ for $0\leq i\leq k$. The coalesced data item has a larger time interval $\tiau=union(\tiau_0,\cdots\tiau_k)$. This reduces the storage overhead. On the other hand, if a data item's $\tiau$ spans across multiple \textit{Time Span Chunks}, then it can be split into multiple data items $\langle e, tp, \tiau_i, v\rangle$ whose  $\tiau_i$ are coalesced to the time interval of the corresponding \textit{Time Span Chunks}.

\subsubsection{Updating the TIM-Tree}
Updates of TIM-Tree arrive in the form of $\langle e, tp, \tiau, v\rangle$. These updates are firstly applied to the in-memory components (\textit{Time Span Chunks}) according to their $tp$ and $\tiau$ (may be split). When an in-memory component reaches a fixed size threshold, it triggers a compaction between disk and memory components, producing a new disk component with merged data and clearing the memory. The new disk component is flushed to disk with bulk sequential writes (rather than random I/O), which improves the overall write efficiency for handling \textit{Append} write in PETG.
The merge process adopts the merge-sort algorithm \cite{CLRS3-MergeSort} whose time complexity is $O(N+m)$ where $N$ and $m$ are the number of data items in the disk and corresponding in-memory component.
% Note that data items whose $\tau_e$ is $Now$ are ,
Additionally, a background process merges small, adjacent \textit{Time Span Chunks} on disk with their in-memory counterparts, achieving the TIM-Tree feature that recent chunks remain small and older ones grow larger.
% and $m$and 
% Updates arriving in the form of $\langle e, tp, \tiau, v\rangle$ are split by $tp$'s \textit{Time Span Chunks} into multiple data items, each applied to the corresponding chunk’s memory component and coalesced with existing items when possible. When a non-$LC$ chunk’s memory component reaches the When $LC$’s memory component reaches the threshold, it spawns a new chunk $NC$; both $NC$ and the smaller $LC$ start with empty memory, as their items are flushed into $NC$’s disk component. 

\subsubsection{Querying the TIM-Tree}
%An Entity-History query $getTP(e^q, tp^q, \tiau^q=[\tau_s^q, \tau_e^q))$ is executed on the TIM-Tree by performing a binary search in memory locating the \textit{Time Span Chunks} matching $tp^q$ and $\tiau^q$, then locating the first data item whose $e=e^q$ and $\tau_s\leq \tau_s^q$, then continuely iterating the following $a$ data items until its $\tau_s\geq \tau_e^q$. The time complexity is $O(log(n)+a)$, where $n$ is the total number of data items of the temporal property and $a$ is the number of items in the query answers.
%For example, consider a scenario where we need to query the history of a specific property (e.g., ``jam status") for an entity (e.g., ``Xue Yuan Road") over a particular time interval (e.g., ``08:00 - 18:00''). The TIM-Tree first locates the \textit{Time Span Chunks} whose property is ``jam status'' and time interval overlaps ``08:00 - 18:00'', then retrieves the historical data from the corresponding \textit{Time Span Chunks} on disk.
An Entity-History query $getTP(e^q, tp^q, \tiau^q=[\tau_s^q, \tau_e^q))$ is answered by:
(1) Binary-searching the in-memory catalogue to locate the \textit{Time Span Chunks} (both on-disk and in-memory) that intersect $tp^q$ and $\tiau^q$.
(2) Within each candidate chunk, locating the first data item satisfying $e = e^q$ and $\tau_s\leq \tau_s^q$ (via binary search).
(3) Sequentially scanning the next item until $\tau_s\geq \tau_e^q$ or $e\neq e_q$, getting $a$ items.
(4) Combining the results from on-disk \textit{Time Span Chunks} with those from their in-memory counterparts. Data items from in-memory components take precedence because they are more recently updated. Since the data volume of an in-memory component is limited and relatively small, this merging step requires only $O(1)$ time.
The overall time complexity is $O(log(n)+a)$ where $n$ is the total number of items for the temporal property and $a$ is the size of the result set in step (3). This approach ensures high performance for Entity-History queries.

\subsection{Transaction Management and Fault Recovery}
%\TGraph supports ACID properties by transaction management and fault recovery. First, it ensures that access to the newly introduced temporal property data is also consistent and compatible with the existing transaction mechanisms of Neo4j. Second, it optimizes the performance of handling typical concurrent read/write workloads of PETG data, reducing the occurrence of deadlocks and waiting situations.
The design goal of transaction management and fault recovery in \TGraph is to ensure the ACID properties of PETG data processing while also providing high transaction performance. However, the transaction mechanism of Neo4j falls short of achieving these objectives once temporal property storage is introduced.
To address these issues, \TGraph extensively refactors the transaction and logging modules of Neo4j. These modules are modified to accommodate the management of temporal property data and are enhanced with a more efficient fine-grained multi-level locking mechanism.
%temporal property data cannot be managed by Neo4j transaction, and the efficiency of the locking mechanism is poor for the data. 
%Consequently, in \TGraph, Neo4j’s transaction and logging modules are extensively refactored: they are generalized to accommodate temporal property data management and furnished with a more efficient fine-grained locking mechanism.

The transaction manager of \TGraph ensures the ACID properties of the access to temporal graph data. The manager runs in two modes: original mode and temporal mode. 
When handling transactions involving no temporal property, the transaction manager runs in its original mode, which falls back to Neo4j's transaction mechanism. 
When handling transactions involving temporal property, the transaction manager would upgrade to its temporal mode, which not only includes all features of the original mode, but also plugins new functions: (1) allocate an in-memory data structure as the transaction's private variable which accepts updates of temporal properties, and (2) create new types of log entries that represent temporal property updates and record them in Neo4j's log. The new functions integrate temporal graph data with Neo4j’s transaction management and fault recovery system, ensuring the ACID properties of \TGraph are fully supported by the modified transaction engine.

%In addition to that, a new locking mechanism is designed and added to the temporal mode of the transaction manager to improve the concurrency performance of the system by reducing conflicts in typical PETG workloads.
The fine-granularity multi-level locking mechanism is designed and added to the temporal mode of the transaction manager to improve the concurrency performance of the system by reducing conflicts in typical PETG workloads.
%We examined the data characteristics of temporal graphs and their concurrent access patterns and found that most requests access only a portion of the time and a portion of the nodes/edges of the entire temporal graph.
Empirical analyses of PETG data and their access patterns show that most requests are limited to specific time intervals and involve only a subset of nodes and edges. However, the current entity-level locking mechanism in Neo4j leads to unnecessary waiting for requests that access disjoint time intervals on the same node or edge, which results in poor concurrency performance. 
To address this issue, we designed a fine-grained multi-level locking mechanism that supports more precise locks and eliminates unnecessary waiting, thus improving overall performance.

%However, current nodes-/edge-level locks in Neo4j can cause unnecessary waiting for requests that conflict in nodes/edges access but do not conflict in time, thus affecting concurrency performance. Furthermore, the number of timestamps in the temporal graph is much larger than the number of its nodes/edges, which worsens the issue. 
%
Table \ref{tab:tg-lock} illustrates the mutual exclusion of the lock mechanism ($\times$ denotes a conflict). 
$L^S(e)$ and $L^X(e)$ are locks at the entity-level on entity $e$ (a node or relationship) and all its properties, where 
$L^S(e)$ is a shared lock and $L^X(e)$ is an exclusive lock. 
$L^S(e, tp, \overline\tau)$ and $L^X(e, tp, \overline\tau)$ are locks at the property-level that can be specific to a particular time interval $\overline\tau$ within the temporal property $tp$ of a given entity $e$. 
By reducing the granularity of locks, we decrease the probability of request conflicts, thereby improving concurrency performance.

\begin{table}[tb]
\centering
\footnotesize
\caption{Mutual exclusion locks in \TGraph}
\vspace{-1ex}
\label{tab:tg-lock}
\begin{tabular}{@{}lcccc@{}}
\toprule
&\hspace{-1ex}$L^S(e)$&\hspace{-1ex}$L^X(e)$\hspace{-1ex}& $L^S(e, tp, \overline\tau_a)$ & $L^X(e, tp, \overline\tau_a)$ \\
\midrule
$L^S(e)$ &  & $\times$ &  &  \\
$L^X(e)$ & $\times$ & $\times$ & $\times$ & $\times$ \\
$L^S(e, tp, \overline\tau_b)$\hspace{-1ex}&  & $\times$ &  & $\hspace{-1ex}\times$ \scriptsize{iff $\overline\tau_a\cap \overline\tau_b\neq\varnothing$}\\
$L^X(e, tp, \overline\tau_b)$\hspace{-1ex}&  & $\times$ & \hspace{-1ex} $\times$ \scriptsize{iff $\overline\tau_a\cap \overline\tau_b\neq\varnothing$}\hspace{-1ex}& \hspace{-1ex}$\times$ \scriptsize{iff $\overline\tau_a\cap \overline\tau_b\neq\varnothing$} \\
\bottomrule
\end{tabular}
\vspace{-2ex}
\end{table}

\section{Implementation}\label{sec:tps-impl}
This section provides details on the implementation of \TGraph, including temporal property storage, transaction management and fault recovery.

\subsection{Temporal Property Storage}
This section presents the implementation of the TIM-Tree, describing its data organization, batch write process, file merge strategy, and data redundancy reduction mechanism.
%\TGraph implements the TIM-Tree using the architecture shown in Fig. \ref{fig:tp-arch}. The nodes in the upper levels of TIM-Tree (above \textit{Time Span Chunks}) are packed in the metadata file of the database and loaded into memory when the system starts. All data written to the module is first buffered in \textit{Time Span Chunks} in memory (called Global/Local Memtables) and subsequently flushed to \textit{Time Span Chunks} on disk (called Stable Files/Unstable Files) and merged using sequential I/Os. This design provides superior write performance for the system. 
%\textit{Time Interval Series} of temporal properties are coalesced during the merge process to save disk space.
\subsubsection{Data organization}
\begin{figure}[tb!]
\centerline{\includegraphics[width=0.95\columnwidth]{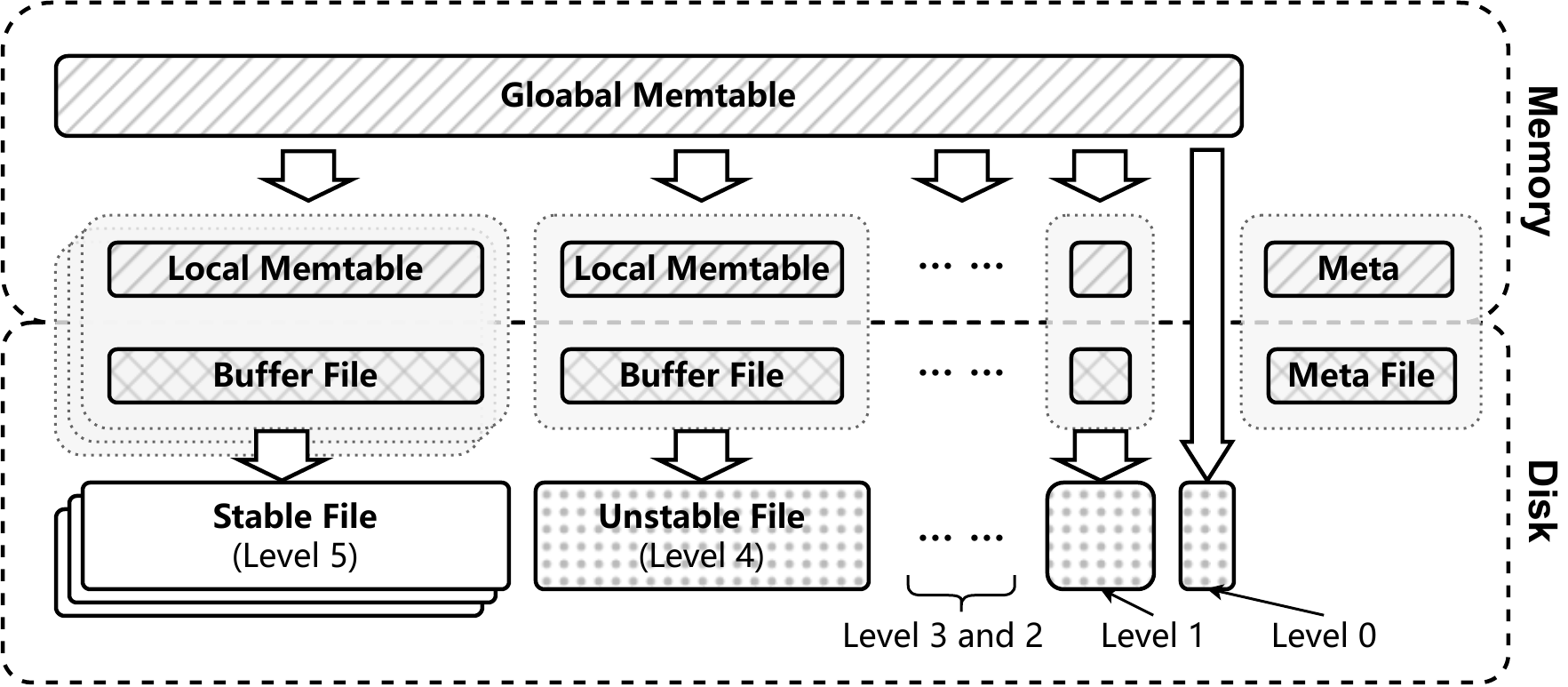}}
\vspace{-1ex}
\caption{Data organization of temporal property storage}
\label{fig:tp-arch}
\vspace{-1ex}
\end{figure}

\TGraph implements the TIM-Tree as shown in Fig. \ref{fig:tp-arch}. The upper-level nodes of the TIM-Tree are stored in the Metadata File and loaded into memory at system startup. 
\textit{Time Span Chunks} on disk are implemented as a series of Stable/Unstable Files, and their in-memory counterparts are implemented as Local Memtables. 
Disk data files utilize SSTables \cite{noauthor_leveldb_2025}, which contain ordered key-value pairs in data blocks and key range indices. 
The Global Memtable (default 64MB, user-configurable), which acts as a write buffer size controller, delegates all updates to Local Memtables.
The Buffer Files serve as backups for Local Memtables during transaction checkpoint events.
% reason for this two-category implementation will be discussed in the file merge strategy section. The disk components of  are Stable/Unstable Files (which are assigned a level from 0 to 5), partitioned by time intervals and grouped by properties using folders.  The in-memory components of \textit{Time Span Chunks} are the Local Memtables, implemented using a Skip-List structure. A Buffer File is used as a backup for each Local Memtable, implemented as logs.
%
% The implementation of $LC$ is a part of the Global Memtable (default 64MB and can be set by users), which is also responsible for allocating updates to Local Memtables.

%The \textit{Time Span Chunks} in memory are implemented using a Skip-List structure. The \textit{Time Span Chunks} on the disk are packed as files and then grouped by properties using folders. The files of each property data are a series of non-overlapping files based on time. This design accelerates queries with a small range of time as querying data within a certain short period of time usually only requires access to one file. 
%Data items on the disk are organized using SSTables \cite{noauthor_leveldb_2025}. An SSTable contains a list of data blocks and an index block; a data block stores key-value pairs ordered by keys, and the index block stores the key ranges of all data blocks.

%\noindent\textbf{Batch writing}: 
\subsubsection{Batch write}
%To minimize the I/O cost of reordering data when modifying disk data, we adopt an "export/merge" approach. All changes to data are first written to the \textit{time span chunks} in-memory (Memtable). A background service periodically merges the data in memory with the data on the disk (using the merge-sort algorithm\cite{CLRS3-MergeSort}) and writes it to the disk in a batch manner. 
To minimize the I/O cost of reordering data when modifying disk data, we adapt an ``export/merge" approach. Changes are first written to the Global Memtable, while a background service periodically merges this data with disk data using the merge-sort algorithm \cite{CLRS3-MergeSort} and flush it to the disk in a batch manner.
Specifically, the service (1) activates when the Global Memtable reaches a set size; (2) allocates updates whose start times before the maximum end time of Stable/Unstable Files to Local Memtables; (3) merges full Local Memtables (default 10MB, user-configurable) to their corresponding Stable/Unstable Files; and (4) packages any remaining updates into a new Unstable File of level 0.
%Memtable is further divided into two layers: Global and Local Memtables. The former temporarily stores all modifications and new data for all time intervals. 
%When the size of the Global Memtable reaches a certain threshold, data beyond the coverage of the existing disk files is exported directly to create new disk files as a level 0 unstable file, while the rest data is transferred to the corresponding Local Memtables. 
%A Local Memtable corresponds to a number of disk files and stores modifications for the data within those files. When the data size of a Local Memtable reaches a certain threshold, it is merged with the disk files. 
%Besides, a Buffer File is created to quickly write Memtable data to disk in response to system log checkpoint events. %, and its content matches that of Memtable.

%\noindent\textbf{File partitioning and upgrade merging}: 
\subsubsection{File merge strategy}
%To reduce write amplification, we optimize storage based on data access patterns. Temporal property data often exhibits write skew, where data closer to the current time is updated more frequently.  %that force Entity-History queries to access a large amount of files
To reduce frequent I/O from small files, a file merge strategy is used. This strategy introduces levels from 0 to 5 to limit unnecessary merging of aged data. Level 5 files are Stable Files with the largest sizes, while Level 0 files are the smallest. Each temporal property can have multiple Level 5 Stable Files and at most one Unstable File in levels 0 to 4. 
After generating a level 0 Unstable File, an extra process checks for existing level n Unstable Files. If another exists, the two files are merged, producing a level n+1 Unstable File (or a Stable File at level 5), and the check is repeated. This strategy maintains the TIM-Tree feature, ensuring smaller recent chunks and larger older ones.
% Here is how the strategy works: 
%To address the write amplification caused by this phenomenon, we partition the disk files into levels L0 through L5 based on the frequency of modifications. 
%L0 files are the smallest in size and store the most recent data, exhibiting the highest update frequency, while L5 files are the largest in size and store the oldest data, characterized by the lowest update frequency.
% \textcolor{red}{Data written to disk right after modification results in smaller storage files, reducing the I/O cost when merging these files with memory data. }
%We develop an ``upgrade merge" mechanism based on the ``export/merge" strategy. This mechanism merges files from adjacent levels to generate higher-level files, such as combining L0 and L1 files to produce L2 files.
%As memory data is continuously written to disk, the related storage files undergo ``upgrade merging". 
%Once files reach the L5 level, they are no longer upgraded further, as these files contain temporal property data that is significantly distant from the current time and relatively stable.

%\noindent\textbf{Reduce data redundancy}: 
\subsubsection{Data redundancy reduction}
To reduce disk data size, we develop three mechanisms: (1) When storing continuously changing property values, we store the data item as $\langle e$, $tp$, $\tau_s$, $v\rangle$ instead of $\langle e$, $tp$, $\tau_s$, $\tau_e$, $v\rangle$, as $\tau_e$ can be calculated from the time of the next preceding data item. This design effectively reduces the disk space usage because most state changes in PETG data occur continuously, allowing us to save space by only storing $\tau_s$. Furthermore, since data items are sorted in ascending order by $\langle e, \tau_s\rangle$ within each file, calculating $\tau_e$ becomes straightforward by accessing the next data item. (2) When a data item's time interval spans multiple files, it is divided into several files for storage. This can be thought of as having a ``checkpoint'' in each disk file. The proportion of stored ``checkpoint'' information to total file data decreases as the size of files increases, which is adjustable through the Global Memtable size set by users. (3) Each data block stored in disk files undergoes prefix compression using the Snappy algorithm \cite{noauthor_snappy_2025}. Since data items in disk files are ordered, there is a high probability of redundancy between adjacent data items, such as identical or similar ids of $e$ and shared high-order bits in $\tau_s$.
% This method typically reduces the storage space needed for temporal properties by more than half.

\subsection{Transaction Management and Fault Recovery}

%The flow of a transaction in \TGraph is illustrated in Fig. \ref{fig:tx-flow}.
%In this section, we explain the transaction management and fault recovery in \TGraph.
In this section, we elaborate on \TGraph’s transaction management and fault recovery module, which extends Neo4j’s native mechanisms while still delegating core ACID guarantees to Neo4j’s original engine. 

Neo4j’s ACID guarantees rest on an in-memory transaction buffer: all writes are staged per transaction and flushed to storage only on successful commit, otherwise discarded, ensuring atomicity and consistency; a write-ahead log (WAL) secures durability, while entity-level locks acquired during updates and released on commit/rollback enforce isolation.

\TGraph extends Neo4j’s transaction pipeline at three points.
Within the in-memory stage, it introduces (i) a structure for temporal data in the transaction buffer, (ii) user-level temporal-graph APIs that inject request data into the buffer, and (iii) a flush routine that writes temporal data from the buffer to storage, leaving Neo4j’s original commit/abort logic intact to guarantee atomicity and consistency.
Within the logging stage: it adds (i) new WAL entry types for temporal data, (ii) a buffer→WAL writer, and (iii) a WAL→storage flusher for crash recovery, all executed under Neo4j’s native logging and recovery mechanisms to ensure durability.
Within the lock-acquiring stage, it employs a fine-granularity multi-level locking mechanism, which acquires an entity-level shared lock first, then a fine-grained temporal exclusive lock for the relevant time interval on write. This eliminates gratuitous waits and improves transaction efficiency. In addition, the fine-grained locks employ Neo4j’s lock and deadlock detection logic in isolated memory, ensuring the correctness of native and temporal locking mechanisms.

\section{Experimental study}\label{sec:exp}
This section evaluate the performance of \TGraph in typical PETG data management scenarios from three aspects: utilization of disk space, latency of read and write operations, and throughput of concurrent transactions. 
% In addition, we discuss the ease of use of \TGraph.

\subsection{Experimental Setup}
% The operating system is Windows 10 Professional (v19044.1645)

\begin{table}[tb!]
\centering
\footnotesize
\caption{Characteristic/statistics of datasets}
\vspace{-1ex}
\label{tab:dataset-stat}
\begin{tabular}{@{}l|ccc@{}}
\toprule
Dataset   & Energy    & Traffic   & SYN   \\
\midrule
Raw data size  & 3.7GB     & 41GB      & 106GB \\
Time span  & 2012-2014 & 2010.5-11 & 1-1B  \\
\# static vertices  & 1.5K      & 80K       & 400K  \\
\# static edges  & 2.2K      & 110K      & 10M   \\
\# events (vertices property update)  & 39M       & -         & 2B    \\
\# events (edges property update) & -         & 1.39B     & 2B    \\
\# temporal properties & V9, E0      & V0, E3      & V3, E1 \\
Temporal property non-redundancy & V0.8 E- & V- E0.4 & V0.3 E0.5\\
\bottomrule
\end{tabular}
\vspace{0ex}
\end{table}

\begin{table*}[t]
\centering
\caption{Details of storage space usage of each system (Bytes)}
\vspace{-1ex}
\label{tab:store-space-result}
\footnotesize
\begin{tabular}{c|cccc|cccc|cccc}
\toprule
Datasets & \multicolumn{4}{c|}{Energy}   & \multicolumn{4}{c|}{Traffic}& \multicolumn{4}{c}{SYN}        \\
Metrics\textbackslash Systems & PETG & PG & MA & Neo & PETG & PG & MA & Neo & PETG & PG & MA & Neo\\
\midrule
Static Data & 2544K & 568K & 496K & 2528K & 38.3M & 29.0M & 25.6M  & 39.2M & 1490M & 1111M & 996M & 1616M \\
%\textcolor{red}{Static Index} & 230K & 167K & 296K & 192K & 13.5M & 7.66M & 12.1M & 10.6M  & 821M & 419M & 519M & 504M \\
Temporal Data & 3.17G & 4.25G & 4.75G & 8.09G & 14.5G & 119G & 133G & 172G & 41.3G & 327G & 365G & 416G \\
%\textcolor{red}{Temporal Index} & 0.00 & 0.04G & 1.15G & 1.86G & 0.00 & 1.56G & 41.0G & 66.0G & 0.00 & 4.54G & 118G & 189G  \\
%\hline
%\textcolor{red}{Total Size (with Index)}    & 3.17G & 7.84G & 4.25G & 4.83G & 14.7G & 118.6G & 119.3G & 136G  & 42.3G & 419G &   328G & 381G \\
Total Size  & 3.17G & 4.25G & 4.75G & 8.10G & 14.6G & 119G  & 133G & 172G & 42.8G & 328G & 366G & 417G \\

Raw Data Size   & \multicolumn{4}{c|}{3.69G} & \multicolumn{4}{c|}{41.1G}  & \multicolumn{4}{c}{106G} \\
\hline
%\textcolor{red}{Ampl. ratio (with Index)}&    \textbf{0.857} & 2.12 & 1.149 & 1.305 & \textbf{0.403} & 2.52 & 2.54 & 3.63   & \textbf{0.402} & 3.99 & 3.12 & 3.63\\
Amplification ratio& \textbf{0.859} & 1.15 & 1.29 & 2.20 & \textbf{0.355} & 2.90 & 3.24 & 4.18   & \textbf{0.404} & 3.09 & 3.45 & 3.93 \\
\bottomrule
\end{tabular}
\vspace{-1ex}
\end{table*}

%\noindent\textbf{Datasets}: 
\subsubsection{Datasets}
Three datasets are used for the evaluation (two real-world temporal graph datasets and a synthetic temporal graph data): (1) Traffic, traffic data of Beijing \cite{temporal_motifs_2025}, (2) Energy, data from European renewable energy power system \cite{jensen_re-europe_2017}, and (3) SYN, data generated by a temporal graph data generator. Some statistical indicators of the datasets are listed in Table \ref{tab:dataset-stat}.
The ``temporal property non-redundancy'' is the average probability that update events of temporal properties in a dataset lead to value changes (``V'' for temporal properties on nodes, ``E'' for those on edges). This metric characterizes how much data can be considered semantically non-redundant.
% 时态属性非冗余率是指数据集中时态属性的更新事件导致状态变化的概率，如果有多个时态属性，则为其平均值。该指标表征了在语义上有多少数据可以被看做是冗余的The number of temporal vertices and edges refers to the quantity of vertices and edges participating in the evolution of temporal properties. For “Number of temporal properties”, “V” represents the count of temporal properties on nodes, and “E” represents that on edges (for instance, “V9, E0” in the Energy dataset indicates there are 9 temporal properties on nodes and 0 on edges). Correspondingly, in the “temporal property non-redundancy rate”, “V” and “E” represent the average probability that update events of temporal properties on nodes and edges, respectively, lead to state changes. This metric characterizes how much data can be considered semantically non-redundant.

%\noindent\textbf{Systems}: 
\subsubsection{Baseline systems}
We compared \TGraph (referred to as PETG) with relational databases (PostgreSQL 17.5, referred to as PG), temporal relational databases (MariaDB 11.4.7, referred to as MA), and graph databases (Neo4j 4.4.37, referred to as Neo). These databases were chosen because they support user-level transactions, a valid-time data model, and are currently common solutions for temporal graph data management. Other temporal graph systems are not considered because they only natively support temporal graph data with transaction-time  \cite{theodorakis_aion_2024, hou_aeong_2024} or do not support full user-level transactions \cite{cheng_kineograph:_2012, han_chronos:_2014, miao_immortalgraph:_2015, khurana_storing_2016, labouseur_g*_2015, byun_chronograph_2020, rost_distributed_2021, massri_clock-g_2022}. 

\subsubsection{Applications and Configurations}
We implement applications on both \TGraph and the baseline systems. These applications support typical queries of PETG data (see Section \ref{sec-query}), including four read operations: \textit{Entity-History}, \textit{Snapshot}, \textit{GATP (max)}, and \textit{ETPC}; two write operations: \textit{Append} and \textit{Update}; and a temporal graph analytical query \textit{Reachable Area} (only available on the Traffic dataset). % that has a temporal property whose values are positive on all edges

When loading data on Neo4j, we create nodes that represent a time interval and link these nodes with entities in the temporal graph \cite{cattuto_time-varying_2013}. 
For PostgreSQL, we create a table to store the nodes with their static properties and a table to store the relationships with their static properties. The temporal properties of nodes/rels are stored in corresponding tables in the form of $\langle e, \tau_s, v_{tp_1}, \cdots, v_{tp_n}\rangle$. 
For MariaDB, valid-time temporal tables are used to store the temporal property data in the form of $\langle e, \tau_s, \tau_e, v_{tp_1}, \cdots, v_{tp_n}\rangle$. 
Note that for PostgreSQL and MariaDB, indexes on $\langle e, \tau_s\rangle$ are created to accelerate the \textit{Entity-History} query. 

The experiments are carried out on a Dell T630 workstation with Intel Xeon(R) CPU E5-2640v3@2.60GHz (x2, 16/32 physical/logical cores) and 192GB memory, with a storage of Dell PERC H730 Adp RAID1 disk array (SAS 7200 rpm, 8TB). 
The operating system is Ubuntu 24.04.2 LTS, with OpenJDK (version 11.0.27+6) installed. For \TGraph and each baseline system, we allocate 48GB memory (\texttt{Xmx=48g} for systems based on JVM) and use the ``read-committed'' isolation level for transactions by default.

\subsection{Space Usage}

% \begin{table}[h]
% \centering
% \caption{Comparison of database disk space usage (GB)}
% \label{tab:exp-space}
% \begin{tabular}{@{}lrrr@{}}
% \hline
%        & Energy & Traffic & SYN \\
% \hline
% Raw data size   & 3.7    & 47      & 105 \\
% PG     & 5.1    & 119     & 328 \\
% MA     & 3.6    & 141     & 383 \\
% Neo4j  & 7.8    & 118     & 418 \\
% TGraph & 3.2    & 15      & 42 \\
% \hline
% \end{tabular}
% \end{table}

\begin{figure}[tb]
\centerline{\includegraphics[width=0.88\columnwidth]{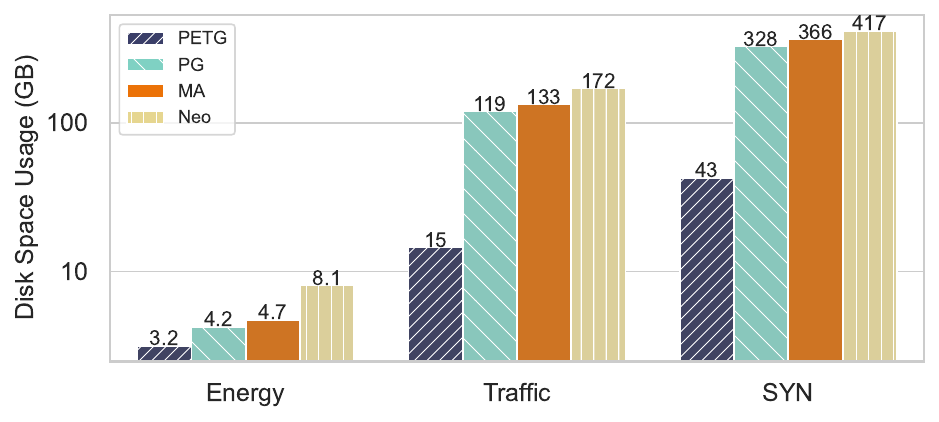}}
\vspace{-1ex}
\caption{Evaluation of database disk space usage (GB).}
\vspace{-1ex}
\label{fig:space-diff}
\end{figure}

%To analyze the space usage of \TGraph and the baseline systems, we store the full datasets Energy, Traffic, and SYN on them and check their disk space usage.
To demonstrate the space efficiency of \TGraph, we analyze the space usage of \TGraph and the baseline systems, store the full datasets Energy, Traffic, and SYN on them, and check their disk space usage (excluding transaction logs).
The results are shown in Fig. \ref{fig:space-diff} and Table \ref{tab:store-space-result}.
% Fig. \ref{fig:space-diff} and 

\TGraph takes up the least storage space (on average 33\% of PostgreSQL, the current best solution) and has more advantages (12.6\% and 13.1\% of PostgreSQL) on the Traffic and SYN datasets, whose non-redundancy ratio is lower than the Energy dataset.
This shows the power of \TGraph's space-saving abilities in data model and temporal property storage.
Table \ref{tab:store-space-result} lists the details of storage space usage, from which we can see the majority of space is occupied by the temporal parts of PETG data, whose space-saving strategies decide the final space cost.
The storage amplification ratio (database space cost to raw data size) is 66.7\% smaller than PostgreSQL on average (25.3\%, 87.7\% and 87.1\% in these datasets, respectively).
% 0.859, 0.355 and 0.404 for \TGraph in datasets Energy, Traffic and SYN, while it is 1.15, 2.90 and 3.09 for 
% and has the smallest storage amplification effect, and the amplification ratio varies in accordance with the temporal property non-redundancy rate of the datasets.
%The size of disk space is 3.17GB, 14.7GB and 42.3GB for \TGraph in datasets Energy, Traffic and SYN, respectively, while it is 4.25GB, 119.3GB and 328GB for PG (the most space-saving existing solution) in these datasets.
%It is even smaller than the raw data size.
%That is, the space cost of \TGraph is average of the best existing solution PG. 
It is worth to point out that the storage space of \TGraph is even smaller than the raw data size, its storage amplification ratio is approximately the same as the datasets' temporal property non-redundancy rate (0.859 vs 0.8 in Energy, 0.355 vs 0.4 in Traffic, 0.404 vs 0.4 in SYN), which indicates that the temporal property storage effectively eliminates redundant information in the temporal data.

%\noindent(2) The temporal property data occupies most of the space in all systems and \TGraph uses the smallest space to save the temporal property data. 
%\textcolor{red}{The size of temporal property data to the total data is 99.95\%, 99.3\% and 95.27\% for \TGraph in datasets Energy, Traffic and SYN, while it is 72.7\%, 65.6\% and 63.7\% for PG in these datasets, respectively.}
%Its \textcolor{red}{ratio to the raw data size} ("TG/Raw", i.e. last row in Table \ref{tab:store-space-result}) is very close to the "temporal property non-redundancy rate" of the dataset. 
%This indicates that the \TGraph temporal property store effectively compresses the redundant information in the temporal data. %, and the compression rate is close to the theoretical maximum value.
%The size of temporal property data (and indexes) is 3.17GB, 14.7GB and 42.3GB for \TGraph in Energy, Traffic and SYN dataset. While this metrics for PG (best existing solution) is 4.25GB, 119.2GB and 327GB in three datasets.

%This part of the \TGraph data is much smaller than that of other comparison systems, even smaller than the raw data size. 

%\subsection{Read and Write}
\subsection{Latency}

\subsubsection{Latency of read and write operations}
To test the latency of \TGraph, we evaluate the response time of read and write operations of \TGraph and the baseline systems with seven typical operations defined in Section \ref{sec-query} on three datasets. 
Each test was performed by randomly generating 10k requests for each type of operation on each dataset. 
%The analysis algorithm is the temporal reachable area of the temporal graph. 
%Single-thread testing
The results are shown in Table \ref{tab:exp-rw} wherein the numerical value represents the 90\% percentile of the response time of these requests, indicating that 90\% of the requests can be completed within this time limit. The empty cell ``-'' represents cases in which the system did not respond to any requests within 1 hour.
%(3.6 million milliseconds, indicating timeouts). 
We find the following:
%No comparison was made on the Energy dataset of Reachable area test\footnote{The temporal reachable area algorithm requires at least one temporal property with a positive value on the edge, but the temporal properties of the Energy dataset are all defined on nodes.}.
% energy           85.8
% ehistory: 0.333  3.0
% snapshot: 0.060  16.7
% gatp(max): 0.021  46.5
% etpc:     0.021  47.7
% append:   1.375  0.727
% update:   0.003  400.0
% traffic          117.6
% ehistory: 0.086  11.7
% snapshot: 0.014  72.9
% gatp(max): 0.023  43.2
% etpc:     0.014  72.8
% append:   0.750  1.333
% update:   0.002  460
% reachable 0.006  161.15
% syn              622.9
% ehistory: 0.023  44.0
% snapshot: 0.461  2.17
% gatp(max): 0.369  2.71
% etpc:     0.358  2.79
% append:   0.005  185.7
% update:   0.000  3500

\begin{table}[tb!]
\centering
\footnotesize
\caption{Comparison of requests latency of read and write operations (milliseconds, 90\% percentile, - for timeout)}
\vspace{-1ex}
\footnotesize
\label{tab:exp-rw}
\begin{tabular}{c|l|cccc}
\toprule
Dataset & Operation & PETG & PG & MA & Neo \\
\midrule
\multirow{6}{*}{Energy} & Entity-History & 2.0 & 6.0 & 64 & 18k \\
 & Update       & 5.0    & 2.0k & 6.4k  & 35k    \\
 & Append       & 11     & 8.0  & 7.1k  & 992k \\
 & Snapshot     & 156    & 2.6k & 76k   & 18k  \\
 & GATP(Max)     & 159    & 7.4k & 17k   & 19k  \\
 & ETPC         & 132    & 6.3k & 73k   & 19k  \\
\hline
\multirow{7}{*}{Traffic}  & Entity-History  & 3.0    & 35  & 344  & 1.2m \\
 & Update           & 5.0   & 2.3k      & 21k   & -  \\
 & Append           & 6.0   & 8.0       & 17k   & -  \\
 & Snapshot         & 5.2k  & 379k      & -   & 1.5m  \\
 & GATP(Max)         & 10k   & 432k      & 873k  & 1.3m  \\
 & ETPC             & 4.6k  & 335k      & -   & 1.0m  \\
 & Reachable Area   & 695   & 112k      & 395k  & -  \\
\hline
\multirow{6}{*}{SYN} & Entity-History  & 1.0 & 44 & 18k  & -   \\
 & Update      & 6.0    & 21k   & 901k  & -     \\
 & Append      & 7.0    & 1.3k  & 101k  & -     \\
 & Snapshot    & 553k   & 2.4m  & 1.2m  & -     \\
 & GATP(Max)    & 443k    & 2.0m  & 1.2m  & -     \\
 & ETPC        & 394k   & 1.1m  & 1.1m  & -     \\
\bottomrule
%OLAP & Reachable Area & \multicolumn{4}{c|}{/}        & 2.6k   & 63k  & 132k & -     & 65     & -    & -    & - \\
%\hline
\end{tabular}
\vspace{0ex}
\end{table}

\noindent (1) \TGraph exhibits a significant overall advantage on latency compared to the current best system (i.e., PostgreSQL and MariaDB) and has more advantage when the dataset gets larger. The operations are speeding up by $267\times$ on average ($85.8\times$, $117.6\times$, and $622.9\times$ for datasets Energy, Traffic, and SYN, respectively). 
Among 19 read/write scenarios, \TGraph outperforms the current best system (PostgreSQL and MariaDB) in 18 scenarios, 
% , speeding up  by 1.4x to 472x
% \textcolor{red}{(summary the overall advantages, how much/quantization, why)} 
and shows a performance improvement of 1 or 2 orders of magnitude in 13 of 19 scenarios.
% Neo4j incurs substantial overhead on read/write performance on temporal data due to its complex modelling, as shown in Table \ref{tab:exp-rw}. We therefore exclude it from the following discussions.

\noindent (2) \TGraph achieves an average improvement of $19.6\times$ on the Entity-History query. This improvement also scales with dataset size ($3.0\times$, $11.7\times$ and $44\times$ for dataset Energy, Traffic and SYN, respectively). 
This improvement is due to \TGraph's design of TIM-Tree partitioning and sorting of temporal property data by time, while MariaDB and Neo4j can only index either the start or end of time intervals, which only partially accelerates the queries, causing all queries and update operations to involve traversing unnecessary time points. For a temporal property with a total of $n$ data items, \TGraph's worst-case complexity of Entity-History query is $O(log(n))$, while other systems can be $O(n)$.
PostgreSQL's performance is closest to that of \TGraph, as it uses B-Tree \cite{btree_survey_1979} indexes, providing a time complexity of $O(log(n))$. However, it requires access to both the index file and the data file, while \TGraph has internal ordering and does not require extra access to the index.
% PostgreSQL的性能是和tgraph最接近的，因为其使用了B-Tree索引提供了和Tgraph相同的时间复杂度，但它需要同时访问索引文件和数据文件，而tgraph内部有序无需访问索引

% ehistory: 1.4    11.7    471.7 = 484.8   /3= 161.6  read 70.57
% snapshot: 9.6    93.4    3.045 = 106.045 /3= 35.34
% gatp(max): 35.3   107.8   4.41  = 147.51  /3= 49.17
% etpc:     12.3   93.9    2.33  = 108.53  /3= 36.18
% append:   0.313  1.63    9.83  = 11.773  /3= 3.92   write 62.36
% update:   220    69.4    72.9  = 362.3   /3= 120.8

% |update-append|
% TG: 11     17     196 
% PG: 1.1k   2.5k   1.1k
% \TGraph exhibits the most advantage on the latency of \textit{Update} operation (compared to the current best system PostgreSQL) 
\noindent(3) The latency of the \textit{Update} operation is significantly larger (at least $16.2\times$) than the \textit{Append} operation for PostgreSQL, 
%while for \TGraph its difference is less than 20 milliseconds. 
while they are close (the difference is less than 20 milliseconds) in \TGraph. 
%This is because PostgreSQL does not need to maintain the ``Point-based Semantics''  \cite{point_based_2004} in the \textit{Append} operation. However, in the \textit{Update} operation, maintaining this semantics requires access (and sort) all temporal property data of the specified entity, which increases its latency significantly. \TGraph's temporal property storage ensures this semantics in each \textit{Time Span Chunk} when creating the chunk, during which only a small subset of the temporal property data of the specified entity is accessed. This results in higher write performance. This semantics is finally implemented during data reading by merge-sort chunks that overlap in time intervals. In this way, \TGraph trades reading performance for writing performance.
This is because PostgreSQL needs to delete items and insert new ones to correctly execute updates, 
with every operation introducing the overhead of maintaining the index, 
while append involves only one insertion with little overhead of index maintenance. \TGraph directly inserts data items into memory and writes them to disk in a batch manner, providing an overall efficiency of data write.
\TGraph is a bit slower than PostgreSQL on the \textit{Append} operation in the Energy dataset because the database for this dataset is small enough to fit into memory and \TGraph involves sort commands in the write operation, while PostgreSQL does not.
% 
% \textcolor{red}{(explain why 2 cases are slower than ...)}
% \noindent(3) \TGraph exhibits a significant advantage. 
% This performance improvement is particularly evident for read queries and analysis algorithms. Regarding writes, other systems introduce additional retrieval, deletion, and update operations to maintain the semantics of temporal property values based on timestamps, leading to higher costs. 

\begin{figure}[tb]
\centerline{\includegraphics[width=0.88\columnwidth]{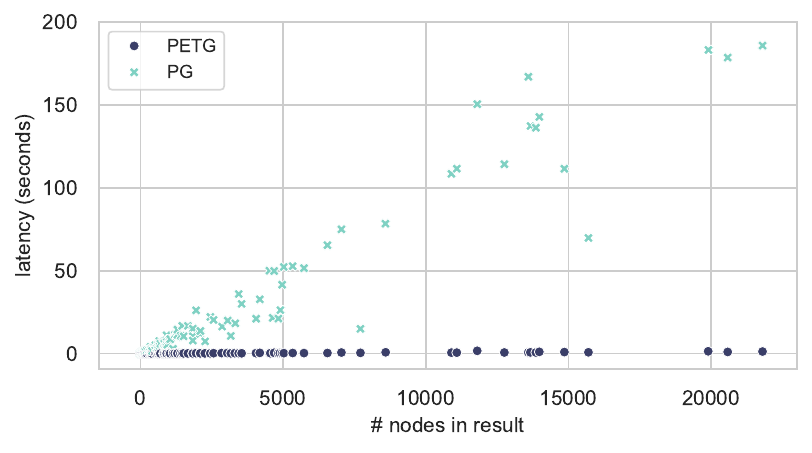}}
\vspace{-1ex}
\caption{Latency of Reachable Area query scales with result size (each data point represents a query)}
\vspace{-1ex}
\label{fig:reachable-area}
\end{figure}

%\begin{figure}[tb]
%\subfigure[getRel Operator]{
%\includegraphics[width=0.48\columnwidth, trim=0.25cm 0.75cm 0.25cm 0.1cm, clip]{figures/reachable-avg_topo_time.pdf}
%}%, trim=0.1cm 0.7cm 0cm 0.2cm, clip
%\subfigure[getTP operator]{
%\includegraphics[width=0.48\columnwidth, trim=0.25cm 0.75cm 0.25cm 0.0cm, clip]{figures/reachable-avg_tp_time.pdf}%, trim=0.1cm 0.7cm 0cm 0.0cm, clip
%}
%\vspace{-2ex}
%\caption{Distribution of single topology and temporal operator latency (milliseconds) in Reachable Area query}
%\vspace{-1ex}
%\label{fig:reachable-area-detail}
%\end{figure}

\begin{figure}[b]
\centerline{\includegraphics[width=0.88\columnwidth]{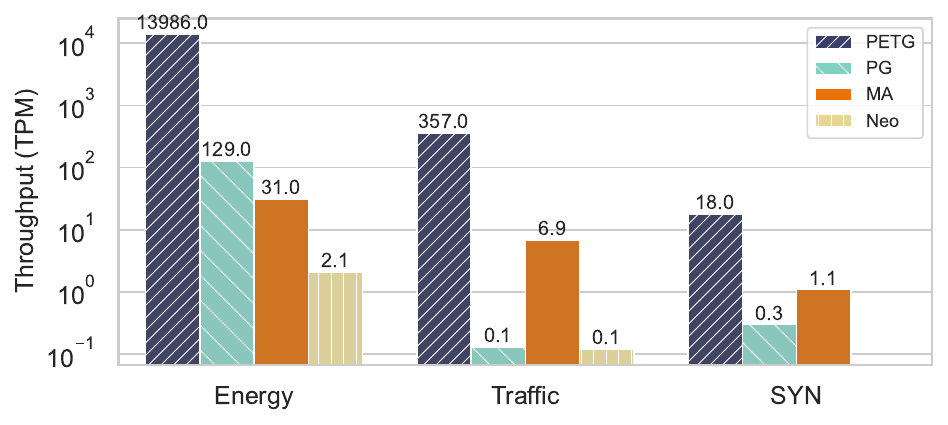}}
\vspace{-1ex}
\caption{Evaluation of transaction throughput on HTAP workload}
%\vspace{-1ex}
\label{fig:parallel-htap-performance}
\end{figure}

% \begin{figure}[b]
% \centering
% \vspace{-1ex}
% \begin{tabular}{cc}
% \subfigure[Throughput (Energy)]{
%     % \label{fig-matrix-a}
%     \includegraphics[width=0.8\columnwidth]{figures/htapThroughput.pdf}
% }% trim=0cm 0.2cm 0.5cm 0.2cm, clip
% \end{tabular}
% \vspace{-2ex}
% \caption{Evalution of transaction throughput}
% \vspace{0ex}
% \end{figure}

\begin{figure*}[tb]
\centering
%\vspace{-1ex}
%\begin{tabular}{cc}
\subfigure[Transaction Throughput]{
    % \label{fig-matrix-b}
    \includegraphics[width=0.62\columnwidth]{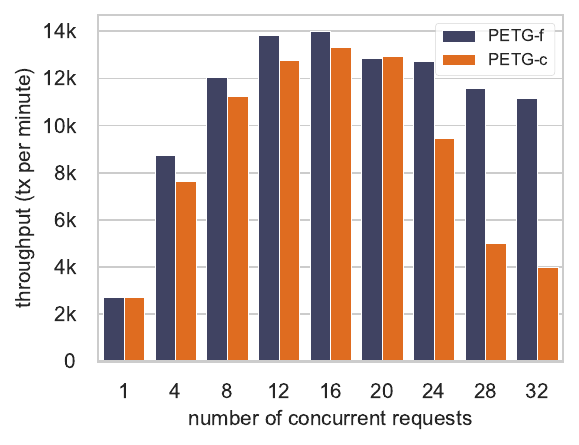}
}
\subfigure[Transaction Latency]{
    % \label{fig-matrix-a}
    \includegraphics[width=0.62\columnwidth]{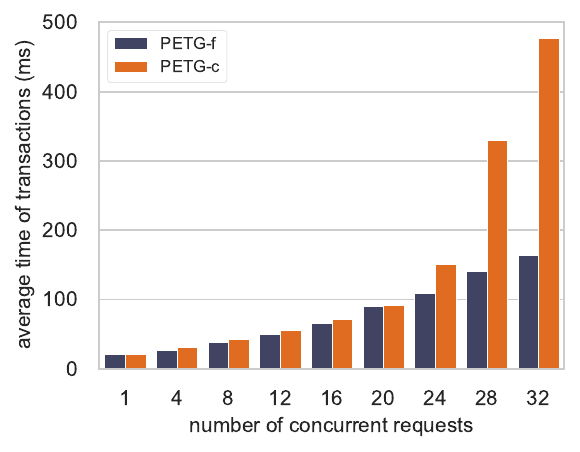}
}
\subfigure[Transaction Dead Lock Count]{
    % \label{fig-matrix-b}
    \includegraphics[width=0.62\columnwidth]{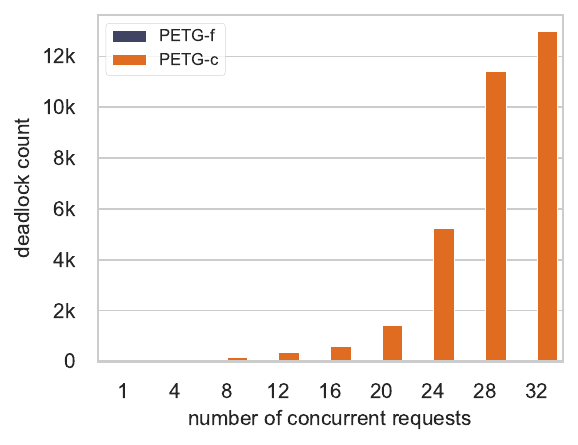}
} \\
%\end{tabular}
\vspace{0ex}
\caption{Effectiveness of fine-granularity multi-level locking mechanism in \TGraph}
\vspace{0ex}
\label{fig:parallel-lock-htap-performance}
\end{figure*}

\noindent (4) \TGraph shows higher speedup (average $34.2\times$) on the analytic queries (\textit{Snapshot}, \textit{GATP(max)}, and \textit{ETPC}) than on the Entity-History query (average $19.6\times$). This greater improvement is caused by the data locality of \TGraph: data items with a closer time interval (of all entities) are packed in adjacent locations on disk, which is not implemented by baseline systems.
% \noindent (4) \TGraph outperforms the current best systems (i.e., PostgreSQL and MariaDB) across the three operations \textit{Snapshot}, \textit{GATP(max)}, and \textit{ETPC}, which, although they are not the primary target of \TGraph. The average improvement is $34.2\times$. This greater improvement over the \textit{Entity-History} query is caused by data locality: \TGraph stores those data items with same property and proximate time interval in adjacent locations on disk, which is not implemented by baseline systems.

\subsubsection{Latency of traversal temporal graphs}
% 我们进一步分析了Reachable area查询中，参数对查询结果的影响，该查询体现了系统处理时态查询和图拓扑查询的综合能力。结果如图3所示，我们发现
To test the ability of \TGraph to simultaneously handle temporal and topological queries, we conduct the following experiment using the \textit{Reachable Area} query (see Algorithm \ref{algo:reachable}) on Traffic, which fetches a sub-temporal-graph whose nodes are reachable within a given time duration departing from the specified node. We generate 100 reachable area queries on the Traffic dataset, randomly selecting a departing node and a maximal travel time range from 120s to 1200s, and track the result size (number of nodes) of the queries. 
The results are shown in Fig. \ref{fig:reachable-area}.

\noindent (1) The execution latency time scales almost linearly according to the number of nodes in the result sub-temporal-graph. 

\noindent (2) \TGraph is on average $119.38\times$ faster than PostgreSQL (the current fastest solution) in queries whose result has at least two nodes. The latencies of \TGraph show a flatter slope, meaning that \TGraph gains more advantage when performing larger queries. 
When fetching a reachable sub-temporal-graph with 19903 nodes, PostgreSQL takes 183.47 seconds, while \TGraph only costs 1.373 seconds ($133.6\times$ faster). 
This advantage is the overall effect of the TIM-Tree data structure of temporal property and the native graph topology implementation of Neo4j. By further extracting the cost of single operators, we observe that TIM-Tree provides a $getTP()$ operator whose average cost is 0.26ms ($23\times$ faster than PostgreSQL), and the native graph store offers a $getRel()$ operator with an average cost of 0.01ms ($70\times$ faster).

\subsection{Transaction Throughput}

\subsubsection{Throughput of HTAP workloads}
To test the transaction throughput of \TGraph, we evaluate the concurrency performance of each system with HTAP workloads on three datasets. Each workload is composed of 100k requests of different types with random parameters. The request pro-portions are: \textit{Append} 40\%, \textit{Update} 5\%, \textit{Entity-History} 35\%, \textit{Snapshot} 5\%, \textit{GATP(max)} 5\%, \textit{ETPC} 5\%, \textit{Reachable} 5\%. 
The maximum number of concurrent connections is set to 16. 
%For each dataset, we generated 10000 requests\footnote{Workload of Energy dataset has 9500 requests with no Reachable requests because the reachable area algorithm requires at least one temporal property with a positive value on all edges, but the temporal properties of the Energy dataset are all defined on nodes.} with random parameters. To simulate the maximum machine hardware utilization in actual data management scenarios, the maximum number of concurrent connections is set to 16. 
The results are shown in Fig. \ref{fig:parallel-htap-performance}.

\noindent (1) The throughput of \TGraph is average 58.8 times higher than the best existing system, i.e., average $108.4\times$, $51.7\times$ and $16.3\times$ on datasets Energy, Traffic, and SYN, respectively. 
This is the combined effect of the technicals applied in \TGraph, especially the TIM-Tree data structure and the fine-granularity multi-level locking mechanism.

\noindent (2) 
The throughput of \TGraph decreases as the dataset size increases. This is because the latency of analytical temporal graph queries (\textit{Snapshot}, \textit{GATP (max)}, and \textit{ETPC}), which access all entities in the graph, is linear with the size of a graph, thus, they cost more time in a larger dataset.

\subsubsection{Effectiveness of fine-grained locks}
To validate the fine-granularity multi-level locking mechanism, we compare fine-grained concurrent locks on HTAP workloads using two types of \TGraph instances: PETG-c (using coarser granularity locks) and PETG-f (fine-grained locks), comparing their transaction throughput, average latency, and the number of deadlocks on the Energy dataset. When writing temporal properties, 
% PETG-c would manually add an Exclusive Lock to the corresponding nodes/edges, while PETG-f would automatically add a Shared Lock to the nodes/edges and then create a smaller granularity temporal property read/write lock corresponding to the query time interval. Each transaction in this experiment is retried at a 50ms interval until committed. 
PETG-c adds an Exclusive Lock to the corresponding nodes/edges manually, while PETG-f automatically applies a Shared Lock and uses smaller granularity locks for the query time interval. Each transaction retries every 50ms until committed.
%In addition, to test the fine-granularity multi-level locking mechanism, the workload must contain a proportion of conflicts. We therefore chose Energy, which provides a measurable conflict rate due to its small graph size.
% In addition, the workload must contain a proportion of conflicts to test the fine-granularity multi-level locking mechanism. Therefore, we use Energy that provides a measurable conflict rate due to its small graph size. Under the HTAP workload generated on Energy, any two transactions conflict with a probability of 2.34\% when using coarse-grained locks, 0\% when using fine-grained locks.
% In addition, the workload must include conflicts to test the multi-level locking mechanism. We use Energy for its measurable conflict rate due to its small graph size. In the HTAP workload on Energy, two transactions conflict with a probability of 2.34\% on entity level and 0\% considering time intervals on entities.
To acheive measurable conflict rate in the workload, we use Energy dataset whose size is smallest among all datasets. The conflict rate of two transactions in the workload is 2.34\% on entity level and 0.000067\% considering time intervals on entities.
% We use an HTAP workload on the two datasets, each containing 10,000 requests, with a ratio of 60\% for \textit{Appends}, 5\% for \textit{Updates}, and 35\% for \textit{Entity History} queries. 
The results are shown in Fig. \ref{fig:parallel-lock-htap-performance}. 

%throughput1-32: f/c - 1:   0.1%,   14.4%,  7.3%,   8.1%,   4.9%,   -0.8%,  34.6%,  131.5%, 180.1%, avg:42.2%
%avg latency1-32: f/c:      1.00,   0.87,   0.90,   0.90,   0.94,   0.99,   0.72,   0.43,   0.34,   avg:0.79
\noindent (1) The overall performance of PETG-f over PETG-c is evident. 
%The transaction throughput of PETG-f is higher than that of PETG-c in almost all cases (max connection count from 1 to 32), while the latency of PETG-f is smaller than that of PETG-c in all cases. PETG-f does not have deadlock in all cases. This indicates the advantage of the fine-granularity multi-level locking mechanism of \TGraph.
The throughput of PETG-f is 42.2\% higher than that of PETG-c on average (0.1\%, 14.4\%, 7.3\%, 8.1\%, 4.9\%, -0.8\%, 34.6\%, 131.5\%, 180.1\%, respectively), while the average latency of PETG-f is $0.79\times$ that of PETG-c on average ($1.00\times$, $0.87\times$, $0.90\times$, $0.90\times$, $0.94\times$, $0.99\times$, $0.72\times$, $0.43\times$, $0.34\times$, respectively). PETG-f does not have deadlock in all cases. This indicates the advantage of the fine-granularity multi-level locking mechanism of \TGraph.

\noindent (2) The performance gap between PETG-f and PETG-c is widened as the number of concurrent requests exceeds 20. The transaction throughput decreases when the number of concurrent requests is larger than the number of physical cores of the machine, indicating the competition for computing resources. However, as the number of concurrent requests continuously increases after 20, the throughput of PETG-c exhibits a faster decline, while its latency and deadlock count exhibit a faster increase simultaneously. This indicates that the competition in transactions of PETG-c is for locks rather than computing resources, which is another evidence of the advantage of the fine-granularity multi-level locking mechanism of \TGraph. %For a certain workload, the higher the degree of concurrency, the greater the possibility of conflicts. The possibility of a deadlock grows exponentially as the number of concurrent processes increases.

\section{Conclusion}
%To address the challenges of managing temporal graph data with evolving property values, 
This paper presents \TGraph, a temporal graph data management system. \TGraph supports a native valid-time temporal property graph model, adapts to the data features and access patterns of the property evolution temporal graphs, and guarantees ACID properties. We extended the existing graph data management system (Neo4j) to support the data model and developed an efficient temporal property storage, transaction management, and a high-level query language. 
Experimental results using three temporal property graph datasets show that \TGraph can effectively and efficiently manage large-scale temporal graph data on a single machine with an average of 67\% disk space-saving. The query latency of \TGraph surpasses the existing temporal graph data management solutions by 267 times on average.
% while simplifying the query code to just half the usual amount. 
Besides, its transaction throughput is 58 times that of the current best solutions. 
This work provides valuable insights into the management of temporal graph data.

\section{AI-Generated Content Acknowledgement}
We improve the readability of the paper text using the functions provided by Grammarly (https://grammarly.com) in the text editor: (1)``Improve it'', (2)``Shorten it'', and (3) correct grammar errors and typos.

% \section*{References}
\bibliographystyle{IEEEtran}
% \bibliography{paper}
\bibliography{references}

\end{document}